\pdfoutput=1

\documentclass[prb,twocolumn,amsmath,amssymb,showpacs]{revtex4-1}

\usepackage{type1cm,eso-pic}

\makeatletter

\AddToShipoutPicture*{\setlength{\@tempdimb}{0.0\paperwidth}\setlength{\@tempdimc}{1.0\paperheight}\setlength{\unitlength}{1pt}\put(\strip@pt\@tempdimb,\strip@pt\@tempdimc){\makebox( 450,-75){{\textcolor[gray]{0.50}{\fontsize{9pt}{9pt}
\selectfont{This is a preprint of the paper published in
\href{http://dx.doi.org/10.1103/PhysRevB.85.035211}
{Phys. Rev. B 85, 035211 (2012).}
}}}
}}}
\makeatother

\usepackage{dcolumn}
\usepackage{bm}
\usepackage[pdftex]{graphicx}
\usepackage{xspace}
\usepackage{lpic}
\usepackage{float}
\usepackage{hyperref}

\newcommand{\AlGaN}{$\text{Al}_x\text{Ga}_{1-x}\text{N}$\xspace}
\newcommand{\AlInN}{$\text{Al}_x\text{In}_{1-x}\text{N}$\xspace}
\newcommand{\GaInN}{$\text{Ga}_x\text{In}_{1-x}\text{N}$\xspace}
\newcommand{\AlGaInN}
    {$\text{Al}_x\text{Ga}_{y}\text{In}_{1-x-y}\text{N}$\xspace}

\newcommand{\etal}{\emph{et al}\xspace}

\newcommand{\sro}[2]{
    \ensuremath{\Gamma^{(#1)}_{\text{#2}}}\xspace
}

\newcommand{\srosym}{
    \ensuremath{\Gamma}\xspace
}

\newcommand{\lro}[1]{
    \ensuremath{{S}_{\text{#1}}}\xspace
}
\newcommand{\lropat}[2]{
    \ensuremath{{S}^{(\text{#1})}_{\text{#2}}}\xspace
}

\newcommand{\con}[1]{
    \ensuremath{c_{\text{#1}}}\xspace
}

\newcommand{\xexti}[1]{
    \ensuremath{x^{(#1)}_{\text{ext}}}\xspace
}

\newcommand{\GaInNcon}[2]
    {$\text{Ga}_{#1}\text{In}_{#2}\text{N}$\xspace}

\newcommand{\dummyBig}{\rule[0ex]{0pt}{4ex}}
\newcommand{\dummySml}{\rule[0ex]{0pt}{3.5ex}}
\newcommand{\dummyBtm}{\rule[-1.5ex]{0pt}{1.5ex}}

\newcommand{\atomeye}{{\sc AtomEye}\xspace}

\begin{document} 

\title{Ordering in ternary nitride semiconducting alloys} 

\author{Micha{\l} {\L}opuszy{\'n}ski}
\affiliation{Interdisciplinary Center for Mathematical and Computational
             Modeling, University of Warsaw, Pawi{\'n}skiego 5A, 02-106 Warsaw,
             Poland }

\author{Jacek A. Majewski}
\affiliation{Institute of Theoretical Physics, Faculty of Physics,
             University of Warsaw, Ho{\.z}a 69, 00-681 Warsaw, Poland}


\date{\today}

\begin{abstract}
\noindent We present a thorough theoretical study of ordering phenomena in
nitride ternary alloys \GaInN, \AlInN, and \AlGaN. Using the Monte Carlo
approach and energetics based on the Keating model we analyze the influence of
various factors on ordering in bulk crystals and epitaxial layers. We
characterize the degree of both short-range order (SRO) and long ranger order
(LRO) for different compositions, temperatures and for substrates associated
with different epitaxial strain. For the description of the SRO, the
Warren-Cowley parameters related to the first four coordination shells are used.
The LRO is detected by means of the introduced sim-LRO parameter, based on
the Bragg-Williams approach. The description of the observed long-range 
ordering patterns and conditions for their occurrence follows.
\end{abstract} 

\maketitle

\section{Introduction \label{sec:Introduction}}
Ordering in semiconducting nitride alloys is recently an intensively
investigated issue. Both experimental as well as theoretical activities deal
with mixtures of AlN, GaN, and InN, which are  important from the application
perspectives such as optoelectronics, high-power and high frequency
electronics, and sensors.  There are three essential issues that trigger such
high research interest.

First, there is much experimental controversy over which ordering patterns can
be obtained in nitrides and under what conditions do they occur. Actually,
various modes of ordering are reported to be observed in the nitride samples.
These include clustering and precipitation, \cite{ElMasry1998,McCluskey1998}
compositional modulation,\cite{Liliental2006,Pakula2006} or uniform random
alloy possibly with some degree of short-range
ordering.\cite{Galtrey2007,Ozdol2010, Humphreys2007} However, many of the
aspects here still remain unclear. For example, recently Galtrey and
coworkers,\cite{Galtrey2007} on the basis of three dimensional  atomic probe
(3DAP) measurements, concluded that there is no evidence of clustering in
\GaInN except for the natural spatial fluctuations of composition.  Based on a
previous report by Smeeton,\cite{Smeeton2003} they suggested that observed
"clusters" could be artifacts related to the radiative damage of the sample
during high electron resolution transmission microscopy (HRTEM) observation.
However, this argumentation met with serious critique from TEM experimental
groups,\cite{Kisielowski2007,Bartel2007} pointing out that the radiative damage
impact can be highly minimized throughout the experiment and that during such
careful measurements clustering of In atoms still remains visible. In the
presence of such ambiguities, modeling could provide a valuable insight into
the nature of ordering occurring in nitrides.

The second reason is related to the fact that ordering influences the electronic
structure and the optical properties of alloys. These properties, including
band gap, carrier localization, mobility, etc., are in turn crucial from the
applications viewpoint, which include laser diodes (LDs), light-emitting diodes
(LEDs), high-electron mobility transistors (HEMTs), sensors, etc. It turns out
that even local short-range ordering significantly influences optical properties
of nitrides as well as other semiconducting systems, see e.g. the work of
Bellaiche \etal. \cite{Bellaiche1998} Of course, global long-range ordering has
also a pronounced impact on optical properties in nitride alloys, as it was
demonstrated, e.g., in Ref. \onlinecite{Dudiy2003}. Recently, also Gorczyca
and coworkers published a series of papers comparing electronic structure of
small supercells calculated assuming either clustered or uniform distribution of
cations. \cite{Gorczyca2009, Gorczyca2009a, Gorczyca2010} Their studies show
significant differences for these extreme distribution cases, for materials
including ternary nitride alloys \GaInN, \AlInN, and \AlInN as well as
selected compositions of quaternaries \AlGaInN.  Yet, another important example
of ordering significance for electronic properties is related to the
luminescence intensity of In containing nitride samples. There are indications
that excitonic recombination occurring at In microclusters could significantly
contribute to the luminescence signal. \cite{Chichibu1997, Chichibu2006}
Therefore, the presence or absence of In clusters should have considerable
impact on efficiency of light-emitting devices.

The third issue has methodological origin. The type and degree of homogeneity
in the considered alloy influences the range of methods that can be applied to
modeling its properties. Many theoretical approaches assume the perfectly
random uncorrelated alloy, characterized by both short- and long-range order
parameters equal to zero. Examples of such methods include virtual crystal
approximation (VCA), coherent potential approximation (CPA), special
quasirandom structures approach (SQS), etc. The question about the presence of
correlations and ordering within nitride alloys is, therefore, also very vital
in terms of accuracy and applicability of various modeling schemes. Moreover,
the detailed knowledge about the ordering patterns could provide more realistic
atomistic configuration for input to semiempirical electronic structure
computation methods such as tight-binding or empirical pseudopotential schemes.
Thus, the overall reliability of nitride alloys modeling can be seriously
improved, once more detailed knowledge about ordering is available.

Keeping in mind the above reasons, we try to shed some light on the ordering
phenomena occurring in nitrides. We carry out the detailed studies of ordering
in different nitride ternary alloys including \GaInN, \AlInN, and \AlGaN. We
verify how various factors could influence the ordering in these systems. The
crucial variables include composition, temperature, and epitaxial strain. Our
simulations are performed under assumption of the thermodynamics equilibrium.
Therefore, we do not directly address the effects specific to the growth
method.  The growth of nitride layers is often performed using methods
operating in non-equilibrium conditions such as molecular beam epitaxy (MBE).
Moreover, usually this epitaxial growth takes place on the surface. Therefore,
the features related to surface ordering and details of growth process could in
principle remain "quenched" in the sample, forming a metastable state.
Phenomena of this kind are not directly included in our modeling. The main
computational tool of our investigation is the static Monte Carlo method. This
is not the only option. There are also kinetic studies available in the
literature for nitrides.  \cite{Ganchenkova2008} The energetics in our
calculations is computed on the basis of the Keating valence force field model
and, for the sake of simplicity, we focus on the zinc-blende structures. The
studies are carried out in the lattice-coherent thermodynamics regime, i.e., we
assume that the alloy constituents form common lattice and can not decay into
separate components, each with its own independent lattice parameter. This
assumption was often not included and there were numerous studies examining the
behavior in the so called lattice-incoherent case. These approaches rely mainly
on the free-energy difference between alloy and binary constituents relaxed to
their own lattice constants. Examples of this type of calculation for ternary
\cite{Ho1996, Adhikari2004, Biswas2008, Teles2000, Teles2002, Caetano2006,
Purton2005, Ferhat2002, Takayama2000, Karpov2004} and for quaternary nitrides
\cite{Adhikari2004a, Takayama2001} are present in the literature. These studies
provided qualitatively similar phase diagrams with unimodal curves separating
the uniform region in the composition-temperature parameter space. Moreover,
they predicted very high critical temperatures $T_{\text{M}}$, above which the
alloy components are fully miscible. For the most analyzed case of the ternary
\GaInN, typical reported values of $T_{M}$ are in the range of
$T_{M}\approx1500$--$2000$~K (see the work of Liu and Zunger \cite{Liu2008} for
a thorough review of different model findings). These high $T_{M}$ values yield
very low estimates for the maximum In composition, which can be incorporated in
\GaInN in typical growth temperatures, before phase separation occurs. On the
other hand, there are experimental findings showing that it is possible to grow
\GaInN samples containing up to $20$--$30$\% of In without triggering
significant clustering, which is in disagreement with aforementioned
predictions of the lattice-incoherent case. Therefore, recently also a
different approach was introduced, \cite{Liu2008, Liu2009, Chan2010} where the
lattice coherence is assumed. It turned out that in this case $T_{\text{M}}$
decreases to values below typical growth temperatures. Instead of the phase
separation, the random alloy phase with some degree of short-range order is
predicted. Therefore, in our study, we focus on lattice-coherent case, which
seems to correspond better to the alloys obtained during epitaxial growth.
Moreover, we also analyze the influence of the stress related to different
epitaxial substrates employed, as this can be an important factor suppressing
or triggering ordering.  \cite{Karpov1998, Teles2000, Teles2002, Karpov2004,
Liu2008}

In order to characterize the degree of ordering we examine equilibrium
structures obtained from our Monte Carlo simulations. To quantify the degree of
short-range order in the generated samples we use the Warren-Cowley family of
parameters. For description and detection of long-range ordering we employ the
approach based on the Bragg-Williams long range order measure. We present a
complete and systematic study of these parameters in the whole composition
range for all ternary combinations \GaInN, \AlInN, and \AlGaN. We also examine
different temperatures, beginning from growth temperature range to higher
values which could correspond to the sample annealing conditions. Moreover, we
examine how the above ordering metrics are influenced by biaxial strain related
to application of a different substrate for growth process.

The paper is organized as follows. We start with a short overview of the short-
and long-range ordering concepts in Sec.~\ref{sec:DescOrder}. The commonly
employed Warren-Cowley and Bragg-Williams parameters are also briefly discussed
there. Next, in Sec.~\ref{sec:CompDetails}, the details of the employed
computational approach are explained. Section \ref{sec:CompOrder} describes how
the order parameters are computed from the Monte Carlo data. In particular, it
introduces the sim-LRO parameter, constructed on the basis of the
Bragg-Williams characteristic. The sim-LRO metric proposed in this work is a
handy indicator of the long-range ordering emergence during Monte Carlo
simulations. The results for the ternary bulk alloys \GaInN, \AlInN, and \AlGaN
are described in Sec.~\ref{sec:OrderTernBulk}. The ordering in biaxially
strained epitaxial \GaInN layers is studied in Sec.~\ref{sec:OrderTernSubs}.
Finally the paper is concluded in Sec.~\ref{sec:Summary}.

\section{Quantitative description of ordering \label{sec:DescOrder}}

The concepts of order and disorder are intuitively easily understood. However,
there exist many ways of converting this intuition to numbers describing the
degree of ordering in crystals. Therefore, to provide a necessary background
and fix the notation, a brief description of the concepts used throughout the
forthcoming analysis of ordering in nitrides is given in this section. For more
thorough insight into the ordering phenomena, see classical monographs of Ziman,
\cite{Ziman1979} Ducastelle, \cite{Ducastelle1991} or a very informative
introductory paper by Klein. \cite{Klein1951}

\subsection{Long-range order vs short-range order}

When analyzing the ordering in alloys, it is important to distinguish between
two different aspects of this phenomena, namely, the short-range order (SRO)
and the long-range order (LRO). The term short-range order is used to describe
the preference of certain types of atoms to reside near each other. This effect
manifests itself in the form of statistical correlations between occurrences of
atomic types on their respective coordination shells. Usually, alloys exhibit
non-zero degree of short-range order, as a result of energetic preference
towards particular atomic arrangements. The long-range order is related to the
development of a global pattern spread throughout the whole crystal. It is
worth underlining that in the temperatures above 0~K the observed pattern is
never perfect and contains some misplaced atoms. Both types of ordering can be
described in a quantitative manner. In the following we recall the definitions
of the commonly used Warren-Cowley SRO parameter and the Bragg-Williams LRO
characteristic.

\subsection{Warren-Cowley short-range order parameter
\label{sec:SROProps}}
One of the most commonly used measures of SRO is the Warren-Cowley family of
parameters \srosym. \cite{Cowley1950} In multicomponent alloys, these SRO
parameters between components A and B are defined on the basis of conditional
probability of finding an A-type atom in the specified coordination shell of 
a B-type atom:

{\footnotesize
\begin{equation}
\sro{i}{\text{AB}} = 1 - 
  \frac
  {P \left ( \left . A\text{ atom in shell }i\text{ of site }j 
   \; \right | \,
   B\text{ atom on site }j \right )}
  {\con{\text{A}}},
  \label{eq:SRODefinition}
\end{equation}}

\noindent{}where $\con{A}$ denotes the concentration of A-type atoms in the
system. It is straightforward to note that the relation $\sro{i}{AB} < 0$
indicates the preference to AB neighborhood on the $i$th coordination shell,
whereas $\sro{i}{AB} > 0$ indicates anti-preference. For the ideal uncorrelated
random alloy, $\sro{i}{AB} = 0$.  It can be also shown that
\cite{Ducastelle1991}
\begin{equation}
\label{eq:SROSymmetry}
\sro{i}{AB}=\sro{i}{BA},
\end{equation} 
and that for each of the components P the following sum rule holds:
\begin{equation}
\label{eq:SROSumRule}
\sum_{B} \con{B} \sro{i}{PB}=0.
\end{equation}
Therefore, for the alloy $\text{A}_x\text{B}_{1-x}$, we can in principle define
four different SRO measures $\srosym_{\text{AA}}$, $\srosym_{\text{AB}}$,
$\srosym_{\text{BA}}$, and $\srosym_{\text{BB}}$. However, because of one
symmetry relation [Eq.~(\ref{eq:SROSymmetry})] and two sum rules
[Eq.~(\ref{eq:SROSumRule})], actually only one SRO parameter is really
independent. In this paper, we use standard convention and focus on
$\srosym_{AB}$. It is worth underlining that, even though technically we deal
with ternary alloys $\text{A}_x\text{B}_{1-x}\text{N}$, they can be described
identically as in the binary case. This is because nitrogen lattice remains
mostly unaffected, as the probability of substitution in the nitrogen site is
much lower than exchanges in cationic lattice because of the energetic reasons.
Therefore, we focus only on the cationic fcc sublattice and investigate in a
great detail order parameters related to the first four coordination shells
within this lattice.

\subsection{Bragg-Williams long-range order parameter
 \label{sec:LROProps}}

The long-range order must be specified with respect to a certain pattern. To
calculate it, e.g., for $\text{A}_{x}\text{B}_{1-x}$ binary alloy we have to
know which sites should be occupied by atoms of type A, type B, etc. Once we
know the spatial pattern (PAT), we can define the Bragg-Williams long-range
order parameter as 
\begin{equation}
    \lropat{PAT}{A} = \frac{f_{A}^{(\text{PAT})}-\con{A}}{1-\con{A}}.
    \label{eq:LRODefinition}
\end{equation}
For the SRO parameter, a key role was played by the conditional probability,
whereas for the LRO measure the most important variable is
$f_{A}^{(\text{PAT})}$. It denotes the fraction of A sites from pattern PAT,
which are actually occupied by A type atoms in the structure under
consideration. Symbol $\con{A}$ stands here for concentration of A atoms in the
sample. If $\lropat{PAT}{A} \approx 1$, it indicates that the location of A
atoms in the structure is similar to pattern PAT. If $\lropat{PAT}{A}=0$, it
means that no significant similarity to PAT in terms of A-atom locations was
detected.  From $\lropat{PAT}{A}<0$, it can be deduced that A atoms in the
structure avoid A sites from PAT. For the binary alloys
$\text{A}_{x}\text{B}_{1-x}$, the $\lropat{PAT}{A}=\lropat{PAT}{B}$, so we have
only one independent Bragg-Williams LRO parameter. 
For the technical reasons, the direct use of the Bragg-Williams
parameter for the quantification of the LRO in Monte Carlo simulation data is
difficult. Therefore, to detect the presence of the LRO in our results, we
develop a derived quantity, the sim-LRO parameter. It is described in 
greater detail in Sec. \ref{sec:CompOrder}.

\section{Computational details \label{sec:CompDetails}}
\begin{figure}
    \includegraphics[width=0.45\textwidth]{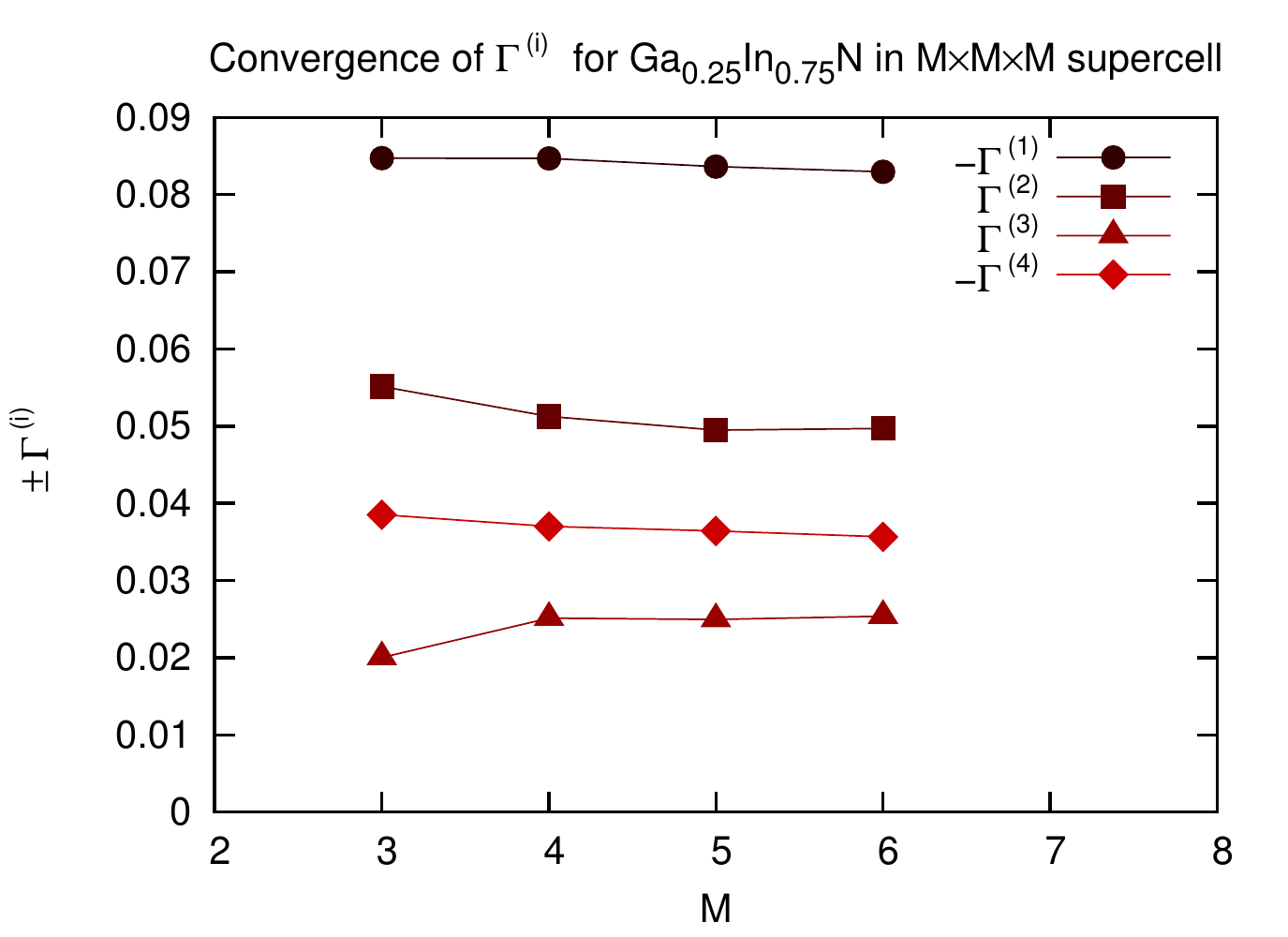}
    \caption{(Color online) Sample convergence test --- the dependence of  the
    Warren-Cowley short-range order parameter \sro{i}{GaIn} on the $M \times
    M \times M$ supercell size in bulk \GaInNcon{0.25}{0.75}. 
    \label{fig:SRO_Convergence}}
\end{figure}

\begin{figure}
    \includegraphics[width=0.48\textwidth]{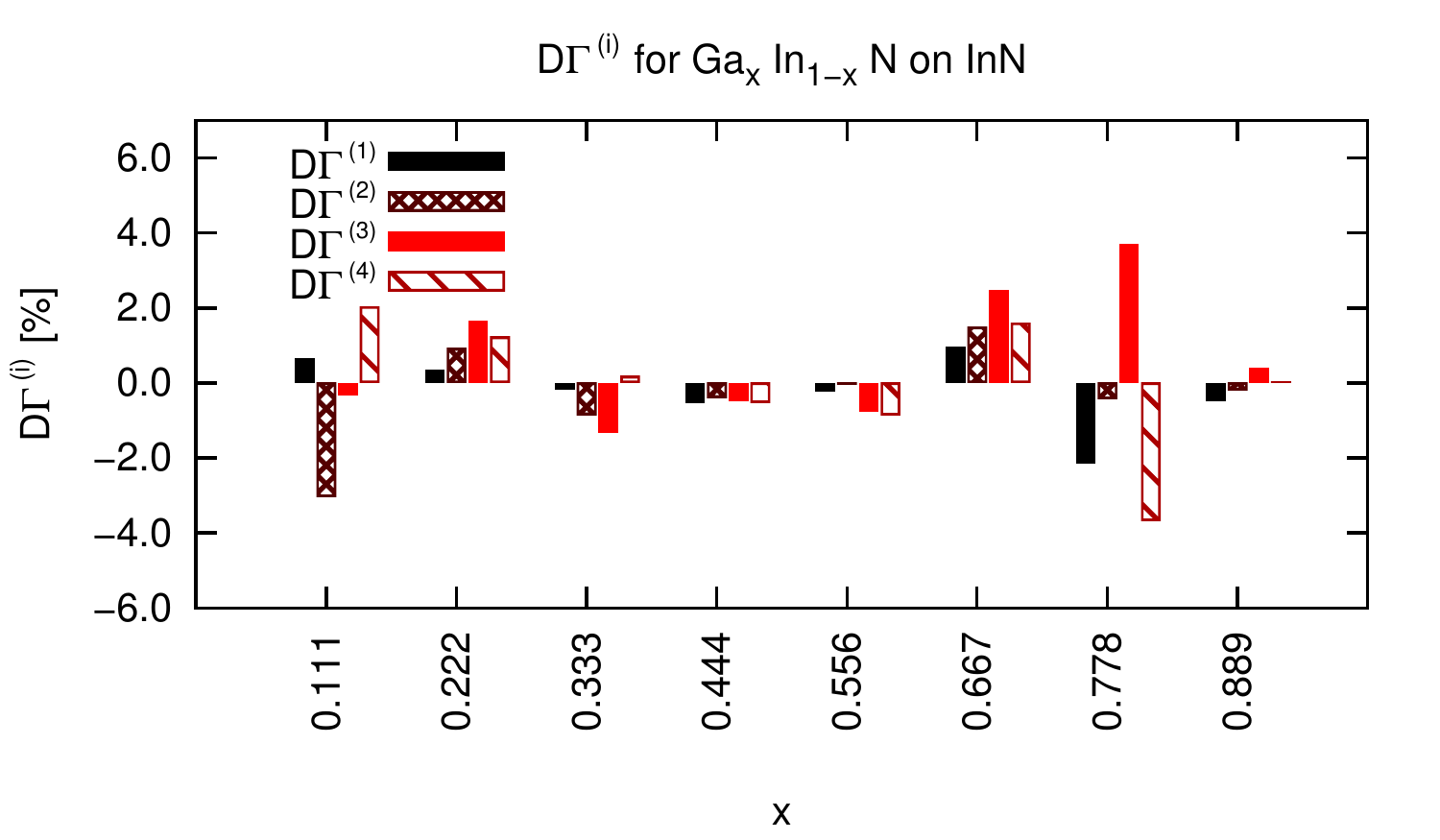}
    \caption{(Color online) Differences between short-range order parameters
    estimation from simulations powered by two
    different random number generator --- RanLux from the GSL library and
    Wichmann-Hill from the ACML package. Simulations were carried out
    for the \GaInN on InN substrate in $T=873$~K.  
    \label{fig:SRO_Gen}}
\end{figure}

All calculations, were performed for the zinc blende $6\times6\times6$ supercell
containing 1728 atoms. As resulted from our tests, this size of the
supercell ensured good convergence of the examined order parameters. See 
Fig.~\ref{fig:SRO_Convergence} for a sample test, showing the dependence of
\sro{i}{GaIn} in bulk \GaInNcon{0.25}{0.75} on the simulation supercell size.
When it comes to the applied simulation techniques, the standard static
Metropolis algorithm within the NVT ensemble was applied. The concentration of
species within alloy was held constant during the simulation, atomic shifts and
cationic exchanges were employed as trial moves. Because of the employment of
atomic shifts the lattice vibrational effects are directly included in the
performed calculations. The magnitude of atomic shifts was adjusted so that
exchange probability was equal to 0.5, as recommended for the most efficient
phase space exploration.\cite{Allen1989,Frenkel2001} The exchange trials were
performed five times more often than atomic shifts, since in the majority of
the cases their probability was much lower than for shifts. The total length of
simulation was 8.3 million Monte Carlo sweeps per concentration. For each run,
we allowed 0.4 million Monte Carlo sweeps for equilibration, before gathering
the simulation statistics. The energy calculations were carried out using
Keating model\cite{Keating1966} and its parametrization for nitrides described
in the earlier work of the authors.\cite{Lopuszynski2010} Periodic boundary
conditions were used throughout all presented simulations. For every type of
ternary alloy, $\text{A}_x \text{B}_{1-x} \text{N}$  a series of simulations
with 17 different concentrations \mbox{$x=$ 0.056, 0.111, 0.167, $\dots$,
0.944} was performed. For the initial conditions, random and uncorrelated
structures were generated with lattice constant optimized to their energy
minimum. For the pseudorandom number generator, the RanLux method was used as
implemented in the GSL library.  \footnote{ \protect
\url{http://www.gnu.org/s/gsl/}} Also tests with different algorithms were
performed, showing that the simulation results are insensitive to pseudorandom
number generation technique employed.  Figure \ref{fig:SRO_Gen} displays the
difference between the \sro{i}{GaIn} parameters computed for \GaInN on InN
using the aforementioned RanLux algorithm and Wichmann-Hill method implemented
in the ACML library, \footnote{\protect
\url{http://developer.amd.com/libraries/acml}} This difference was calculated
according to the formula
{\small
\begin{equation}
D \sro{i}{GaIn} =  \sro{i}{GaIn}(\text{RanLux})
                  -\sro{i}{GaIn}(\text{Wichmann-Hill}).
\end{equation}}
The observed values of $D \sro{i}{GaIn}$ are small. As it can be seen from
Fig.~\ref{fig:SRO_Gen}, they do not exceed a few percent.

\section{Determinations of order parameters from simulation data
              \label{sec:CompOrder}}

The computation of Warren-Cowley parameters \sro{i}{AB} from structures
obtained in Monte Carlo simulations is relatively easy. It is enough to
estimate the conditional probability from formula (\ref{eq:SRODefinition}) by
average percentage of type A atoms present in the $i$th shell of the B atoms in
the set of structures generated during each Monte Carlo run. However, as
opposed to the short-range order, the situation for the long-range ordering is
not so straightforward. To be able to estimate the Bragg-Williams measure
\lropat{PAT}{A}, the pattern PAT has to be known in advance. Its correct
selection is crucial for the detection of the long-range order.  When
calculating \lropat{PAT}{A} for a highly ordered structure, however, very
different from assumed PAT, one may easily obtain values close to zero,
indicating the absence of LRO. Moreover, even if the exact pattern is known,
one should check for all its symmetry-equivalent variations, since the value of
\lropat{PAT}{A} clearly depends on assumed orientation of PAT. Therefore, in
principle, all variants of PAT produced by symmetry operations should be
checked, as the variant developed during simulation is usually a result of
random fluctuations and can not be predicted \emph{a priori}. However, it would
be very useful to have a quick way of checking whether long-range order of some
type develops during Monte Carlo simulation or not, preferably without the
\emph{a priori} knowledge about ordering pattern.

\begin{figure}[!htb]
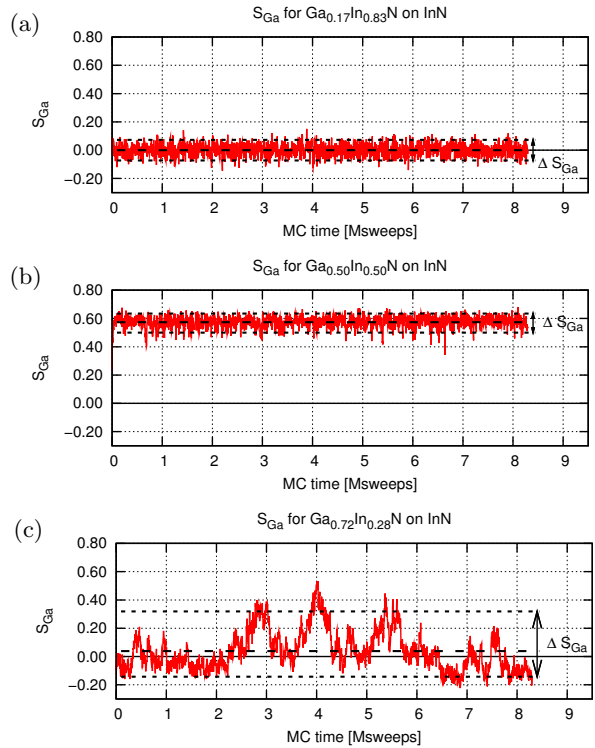

\centering
    \begin{lpic}{lro_Al_000_Ga_288_In_576(8cm)}
    \lbl[l]{0,55;(a)} 
    \end{lpic}
    \begin{lpic}{lro_Al_000_Ga_432_In_432(8cm)}
    \lbl[l]{0,55;(b)} 
    \end{lpic}
    \begin{lpic}{lro_Al_000_Ga_624_In_240(8cm)}
    \lbl[l]{0,55;(c)} 
    \end{lpic}
    \caption{(Color online) Different types of behavior of \lro{Ga} during
             simulations for for \GaInN alloy on InN substrate. Three gallium
             concentrations $x$ are presented: $x=0.17$ (a), $x=0.50$ (b),
             $x=0.72$ (c). Dashed lines correspond to the average value of
             \lro{Ga} and its spread $\Delta \lro{Ga}$.
             \label{fig:LRO_TimeEvolution}}
\end{figure}

It turns out that, in order to detect the presence of long-range ordering,
quantities derived from \lropat{PAT}{A} are more useful. Instead of \emph{a
priori} selecting ordered configuration PAT, it is helpful to calculate the
Bragg-Williams parameter with one of the structures obtained in the developed
stage of simulation used as pattern PAT. It is important that this reference
structure PAT is selected far after the thermalization phase. The quantity
calculated with this method we will call the simulation LRO parameter
(\mbox{sim-LRO}) and denote it as \lro{A}. The typical behavior patterns of
\lro{A} during Monte Carlo simulation run are displayed in
Fig.~\ref{fig:LRO_TimeEvolution}. They are taken from calculations for \GaInN
on InN substrate, which are discussed in greater detail in
Sec.~\ref{sec:OrderTernSubs}. It turns out that three types of behavior for
\lro{A} can be distinguished. First, if no LRO is present then \lro{A} simply
fluctuates around 0 value, corresponding to unordered alloy. This is
illustrated in Fig.~\ref{fig:LRO_TimeEvolution}~(a). Second, if certain LRO
pattern was developed during simulation, the successive structures fluctuate
around it generating values of \lro{A} that are considerably different from 0.
This is depicted in Fig.~\ref{fig:LRO_TimeEvolution}~(b). Third, sometimes
different arrangements of atoms could repeat during simulations, e.g., related
to different orientation or shift of LRO pattern. It will correspond to series
of peaks visible in \lro{A} values during simulation time, see
Fig.~\ref{fig:LRO_TimeEvolution}~(c). These peaks correspond to moments when
the structure is similar to the one selected for comparison. Even though, the
average values of \lro{A} throughout the whole simulation could be in this case
close to 0, this behavior also indicates the development of LRO or
precipitation. Therefore, to detect this type of ordering, except the average
value of \lro{A} we also analyze the spread of observed \lro{A} values.
Convenient measure here is the difference of percentiles $q$ related to 5 and
95 percent, estimated on the basis of \lro{A} population generated during Monte
Carlo (MC) sampling
\begin{equation}
    \Delta \lro{A} = q_{0.95}(\lro{A}) - q_{0.05}(\lro{A}).
\end{equation}
This quantity provides the interval width which includes 90\% of the observed
data. The occurrence of high values for either average \mbox{sim-LRO} parameter
\lro{A} or its spread $\Delta \lro{A}$, enables for convenient detection of long
range ordering in MC simulation results. The average value of \lro{A} together
with the percentiles forming $\Delta \lro{A}$ are marked with dashed lines in
Figs.~\ref{fig:LRO_TimeEvolution}~(a)--(c).

\section{Ordering in the ternary bulk alloys \label{sec:OrderTernBulk}}

In this section, the ordering in the bulk nitride ternary materials \GaInN,
\AlInN, and \AlGaN is examined in a great detail. In the analyzed cases, no
long-range ordering was observed. The \mbox{sim-LRO} parameter for all
presented cases fluctuated around zero, similarly to the situation displayed in
Fig.~\ref{fig:LRO_TimeEvolution}~(a). Therefore, we focus here on the
short-range ordering. The Warren-Cowley SRO parameters for the first four
coordination shells in the cationic lattice are examined in the whole
concentration range.  Moreover, this study covers a few different temperatures,
beginning roughly from the lower end of growth temperatures range, i.e., 873~K,
and including higher values up to 1673~K. This upper limit could correspond to
annealing during the post-processing phase. In the following the results for
each of the ternary combinations, namely, \GaInN, \AlInN, and \AlGaN, are
described.

\subsection{Ordering in \GaInN}

\begin{figure}[H]
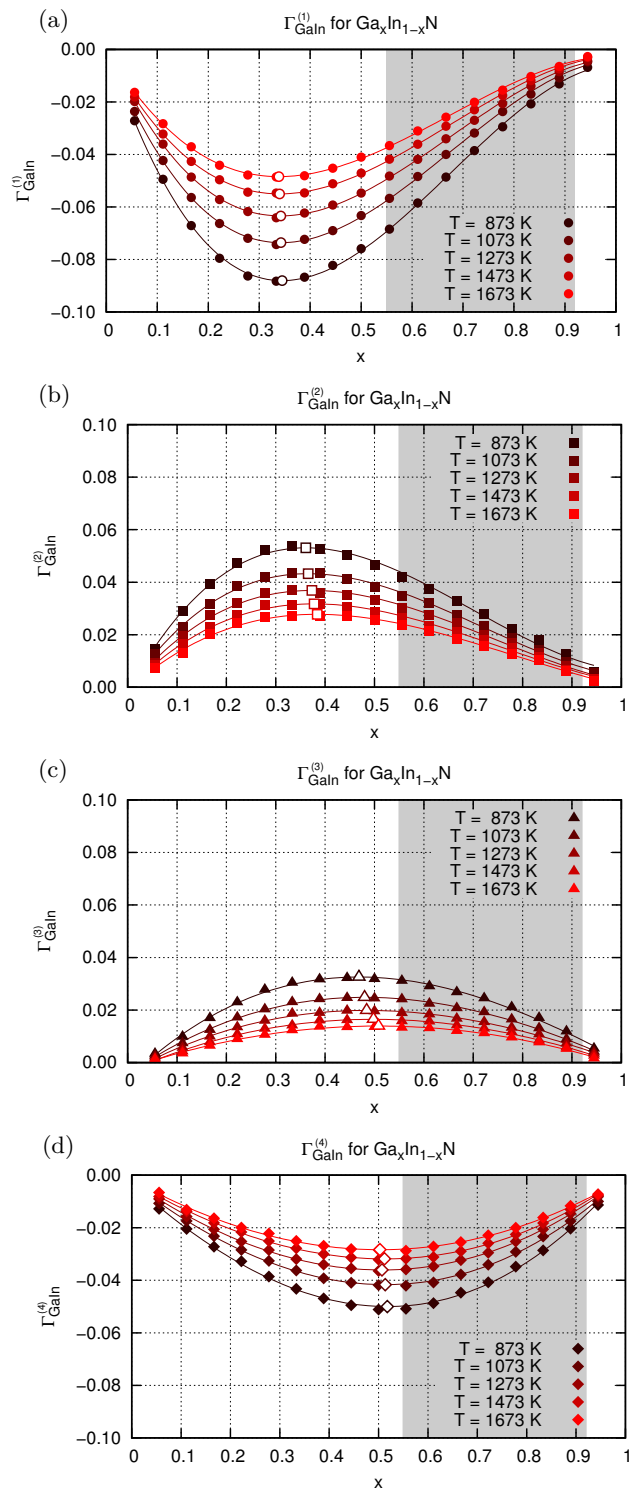

\centering
    \begin{lpic}{oto1_c_GaInN(8.5cm)}
    \lbl[l]{7,83;(a)} 
    \end{lpic}
    \begin{lpic}{oto2_c_GaInN(8.5cm)}
    \lbl[l]{7,83;(b)} 
    \end{lpic}
    \begin{lpic}{oto3_c_GaInN(8.5cm)}
    \lbl[l]{7,83;(c)} 
    \end{lpic}
    \begin{lpic}{oto4_c_GaInN(8.5cm)}
    \lbl[l]{7,83;(d)} 
    \end{lpic}
    \caption{(Color online) The short-range order parameters in \GaInN as a
             function of composition for different temperatures. Note that the
             scale on all graphs is the same to allow for direct comparison of
             ordering magnitude in different coordination shells. Continuous
             lines represent third order polynomial fits to simulated data. The
             extrema for presented fits are marked with open symbols. The
             composition range approximately corresponding to violet-blue-green
             wavelength of emitted light (400--570 nm) is gray shaded.
             \label{fig:SRO_GaInN}}
\end{figure}

\GaInN is up to now the most investigated of all ternary combinations of
nitrides. Currently, it is the main ingredient of the active region in
blue-green optoelectronic devices. The dependencies of $\sro{1}{GaIn}$,
$\sro{2}{GaIn}$, $\sro{3}{GaIn}$, and $\sro{4}{GaIn}$ on composition for
different temperatures are presented in Fig.~\ref{fig:SRO_GaInN}. The most
interesting composition range, corresponding to violet-blue-green wavelength
400--570 nm is gray shaded on all graphs. It is easy to observe that
\sro{1}{GaIn} and \sro{4}{GaIn} are negative in the whole concentration range
and for all examined temperatures.  This indicates the preference toward Ga--In
neighboring on the first and fourth coordination shells. Conversely,
\sro{2}{GaIn} and \sro{3}{GaIn} remain positive, which indicates that Ga atom
is more likely to have another Ga atom on its second and third coordination
shell, than it would result from concentration $x$. This sign pattern is in
qualitative agreement with recent results of Chan and coworkers \cite{Chan2010}
obtained on the basis of the cluster expansion model.  This $-/+/+/-$ sequence
observed for the \sro{i}{} parameters is usually interpreted as manifestation
of stability for the chalcopyrite structure, which exhibits the same SRO signs
pattern.\cite{Chan2010,Chen2008}

For each dependence of the short-range order parameter on gallium concentration
$x$, the third order polynomial was fitted. This type of curve is selected as
the simplest model capable of describing observed shapes with asymmetric
maximum/minimum. These fits are presented as continuous lines in
Fig.~\ref{fig:SRO_GaInN}. It turns out that all short-range order parameters
dependencies have single extremum, corresponding to maximum absolute value of
\srosym parameter. On the basis of obtained polynomial fits, concentrations
\xexti{i} corresponding to this maximum ordering are determined. These extrema
are marked with open symbols in Fig.~\ref{fig:SRO_GaInN}. The numerical values
of obtained \xexti{i} are presented in Table~\ref{tab:GaInNExtrema}. For the
nearest neighbors and next nearest neighbors, the \xexti{i} is located at lower
gallium concentration roughly around 35\%, or in other words at high indium
content 65\%. These compositions are outside of the most interesting
violet-blue-green range. For the third and fourth coordination shells, these
extrema move toward approximately 50\% compositions. Also the overall curve
shape gets more symmetric around extremum for the case of \sro{3}{GaIn} and
\sro{4}{GaIn} than it is in the case of the first and second shell.
Nevertheless, these higher-order parameters should have weaker influence, e.g.,
on electronic properties of \GaInN than \sro{1}{GaIn} and \sro{2}{GaIn}.
Therefore, we conclude that the region, where effects related to ordering are
the most pronounced is located around $x \approx$ 0.35.
It is also worth noting that the position of $\xexti{i}$ is virtually
independent on temperature, and the difference
\begin{equation}
\label{eq:DeltaCon}
    \Delta \xexti{i} = 
        \xexti{i}(T=1673K)-\xexti{i}(T=873K)
\end{equation}
for all coordination shells $i=1,\dots4$ does not exceed 0.04.
We also calculate the difference of extremum values for each \sro{i}{}
\begin{equation}
    \label{eq:DeltaSRO}
    \Delta \sro{i}{ext} = 
        \sro{i}{ext}(T=1673K)-\sro{i}{ext}(T=873K).
\end{equation}
This quantity indicates that the ordering decreases roughly by half when 
the temperature changes from $T=873$ to $T=1673$~K. All the above findings are
summarized in Table~\ref{tab:GaInNExtrema}.

As far as modeling of electronic structure for \GaInN  is concerned, the above
results indicate that it would be the most interesting to focus on structures
with \sro{i}{GaIn} having successive signs $-/+/+/-$ as an input to density
functional theory, tight-binding or empirical pseudopotential schemes. Methods
which assume 
\noindent $\sro{i}{GaIn}=0$ such as, e.g., special quasirandom structures
(SQS) or virtual crystal approximation (VCA) could lead to systematic
inaccuracy, caused by neglecting of the SRO. The range, where these inaccuracies
should be the most pronounced, is determined by extrema for the
lowest
coordination shells SRO parameter, namely around indium concentration of 65\% or
so. On the other hand, when considering alloys corresponding to the violet end
of spectra, say around 10\% of indium, the influence of ordering effects should
be much less pronounced.
\begin{table}[!b]
\caption{
The summary of extrema for $\srosym(x)$ dependencies in \GaInN presented in
Fig.~\ref{fig:SRO_GaInN}. Concentrations of maximum ordering $\xexti{i}$,
together with corresponding extremum value of \sro{i}{GaIn} at the lowest
examined temperature $T=873$~K are provided. The influence of temperature is
illustrated using the difference $\Delta \xexti{i}$ between extremum
concentrations at $T=1673$~K and $T=873$~K [see Eq.~(\ref{eq:DeltaCon})]
and extremum values of SRO parameters $\Delta \sro{i}{ext}$ [see
Eq.~(\ref{eq:DeltaSRO})].
\label{tab:GaInNExtrema}}
\smallskip
\begin{ruledtabular}
\begin{tabular}{cdddd}
\dummyBtm &
\multicolumn{1}{c}{$ \xexti{i}$} &
\multicolumn{1}{c}{$\sro{i}{ext}$(T=873 K)} &
\multicolumn{1}{c}{$\Delta \xexti{i}$(T=873 K)}     &
\multicolumn{1}{c}{$\Delta \sro{i}{ext}$} \\
\hline
\dummyBig\sro{1}{GaIn} &   0.35 &     -0.088      &     -0.006     &      0.040\\
\dummySml\sro{2}{GaIn} &   0.36 &      0.053      &      0.023     &     -0.025\\
\dummySml\sro{3}{GaIn} &   0.47 &      0.033      &      0.040     &     -0.019\\
\dummySml\sro{4}{GaIn} &   0.52 &     -0.050      &     -0.015     &      0.022\\
\end{tabular}
\end{ruledtabular}
\end{table}
The comparison of the above results with experiment is not straightforward.
There is a heated debate about the range of concentrations and, more generally,
conditions that lead to In clustering in \GaInN samples. This issue so far is
by no means clarified, even on the phenomenological level. Our computations, to
a certain extent, support these findings that report no clustering or 
long-range ordering in the samples even with high In concentration.
\cite{Galtrey2007,Humphreys2007,Ozdol2010} In our simulations we show that
under the lattice-coherent assumption, thermodynamics does not prohibit
obtaining mixtures without precipitates or LRO in very broad indium
concentration range and in temperatures higher or in the region of the typical
growth conditions. However, one has to stress the fact, that this state could
be difficult to obtain experimentally, mostly due to artifacts related to
epitaxial growth methods. These methods, such as MBE or 
metal-organic vapor phase epitaxy (MOVPE), take place on
the surface, in conditions at least partially corresponding to non-equilibrium.
Therefore, some features specific to growth process could form metastable state
and remain "quenched" in the samples. To fully describe these phenomena, the
detailed growth models should be developed.  This is, however, far more
difficult than already not-easy computation of equilibrium properties. Yet
another issue not addressed on this stage of modeling is the presence of 
strain, both local and global. By local strain we mean deformations related to
defects, in particular dislocations, whereas global strain occurs in thin
layers and is related to lattice mismatch between substrate and alloy material.
We reference it as global, since it is present in the whole volume of the
considered layer. The influence of dislocations and other sources of local
strain is not included in our modeling, since we assume perfect crystalline
lattice. Taking into account such extended defects would require calculations
with much larger supercells with complicated geometry, prohibitive for the
detailed composition and temperature scans provided in this study.
Nevertheless, we can gain certain insight into the influence of strain on
ordering by modeling global epitaxial deformation. It was already indicated in
the literature that this epitaxial strain can have significant impact on
ordering phenomena.\cite{Karpov1998, Teles2000, Teles2002, Karpov2004, Liu2008}
Therefore, its role is analyzed in detail in Sec.~\ref{sec:OrderTernSubs}.
\subsection{Ordering in \AlInN}
When speaking of ternary nitride alloys with application prospects, \GaInN is
not the only option. Recently, also \AlInN attracts considerable attention.  It
is particularly interesting, because by adjusting the In composition one can
lattice match it to GaN substrate, producing in principle strain-free
interfaces and heterojunctions.  The absence of strain has many advantages such
as lack of electric field component related to piezoelectric effect, lower
density of defects on the interface, etc. Therefore, such \AlInN layers are
promising candidates for applications in optoelectronic devices, e.g., for high
quality factor microcavities or distributed Bragg reflectors. Moreover, the
AlInN/GaN junction could be an important building block for high electron 
mobility transistors (HEMTs). For more detailed information about properties 
and applications of \AlInN, see a recent review paper by Butt\'e and 
co-workers. \cite{Butte2007}

\begin{figure}[H]
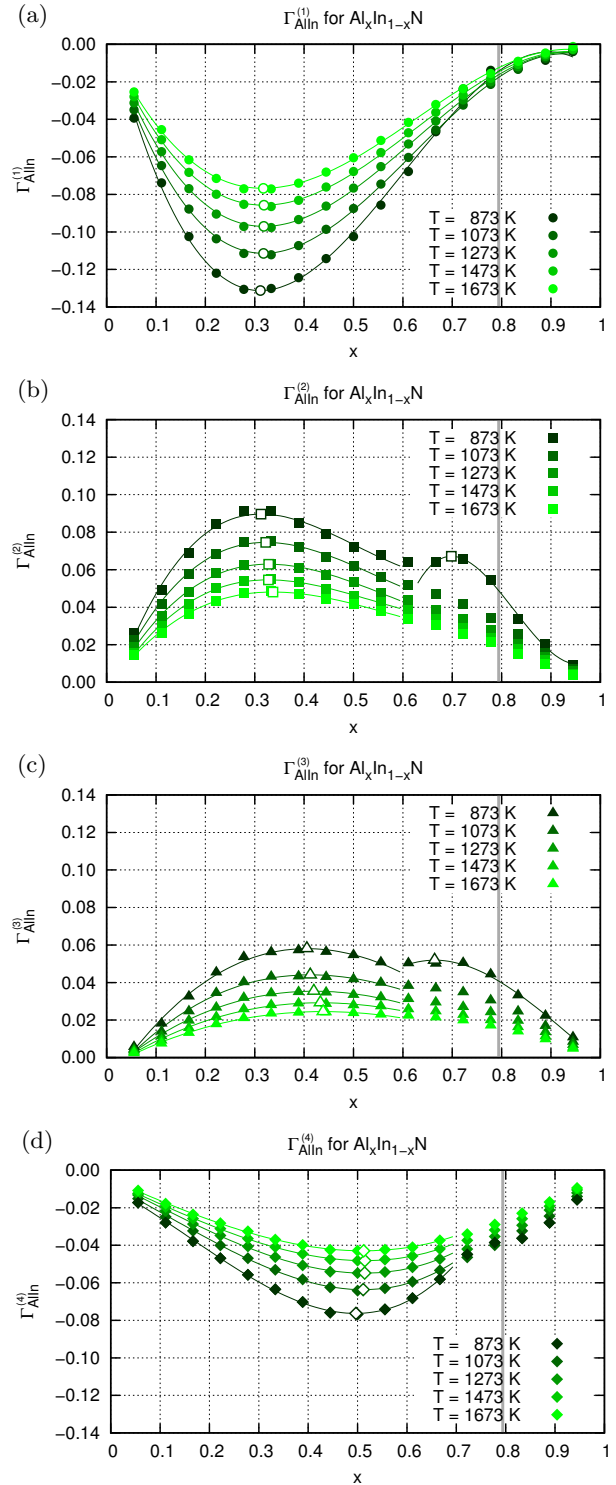

\centering
    \begin{lpic}{oto1_c_AlInN(8.5cm)}
    \lbl[l]{7,83;(a)} 
    \end{lpic}
    \begin{lpic}{oto2_c_AlInN(8.5cm)}
    \lbl[l]{7,83;(b)} 
    \end{lpic}
    \begin{lpic}{oto3_c_AlInN(8.5cm)}
    \lbl[l]{7,83;(c)} 
    \end{lpic}
    \begin{lpic}{oto4_c_AlInN(8.5cm)}
    \lbl[l]{7,83;(d)} 
    \end{lpic}
    \caption{(Color online) The short-range order parameters in \AlInN as a
             function of composition for different temperatures. Note that the
             scale on all graphs is the same to allow for direct comparison of
             ordering magnitude in different coordination shells. Continuous
             lines represent third order polynomial fits to simulated data. The
             extrema for presented fits are marked with open symbols. The
             composition of \AlInN lattice matched to GaN is indicated with gray
             line. \label{fig:SRO_AlInN}}
\end{figure}

\begin{table}[!bht]
\caption{
The summary of extrema for $\srosym(x)$ dependencies in \AlInN presented in
Fig.~\ref{fig:SRO_AlInN}. Concentrations of maximum ordering $\xexti{i}$,
together with corresponding extremum value of \sro{i}{AlIn} at the lowest
examined temperature $T=873$~K are provided. Since in \AlInN, for 
\sro{2}{AlIn} and \sro{3}{AlIn} two extrema are present, both of these are 
quoted in the table. The influence of temperature is
illustrated using the difference $\Delta \xexti{i}$ between extremum
concentrations at $T=1673$~K and $T=873$~K, [see Eq.~(\ref{eq:DeltaCon})]
and extremum values of SRO parameters $\Delta \sro{i}{ext}$ [see
Eq.~(\ref{eq:DeltaSRO})].
\label{tab:AlInNExtrema}}

\smallskip
\begin{ruledtabular}
\begin{tabular}{cdddd}
\dummyBtm     &
\multicolumn{1}{c}{$\xexti{i}$} & 
\multicolumn{1}{c}{$\sro{i}{\text{ext}}$(T=873 K)} &
\multicolumn{1}{c}{$\Delta x_{\text{ext}}$(T=873 K)} &
\multicolumn{1}{c}{$\Delta \sro{i}{ext}$} \\
\hline
\dummyBig$\sro{1}{AlIn}$ &   0.31 &     -0.131      &      0.006     &      0.055\\
\dummySml$\sro{2}{AlIn}$ &   0.31 &      0.090      &      0.024     &     -0.042\\
\dummySml                &   0.70 &      0.067      &                \\
\dummySml$\sro{3}{AlIn}$ &   0.41 &      0.058      &      0.033     &     -0.034\\
\dummySml                &   0.66 &      0.052      &                \\
\dummySml$\sro{4}{AlIn}$ &   0.50 &     -0.076      &      0.016     &      0.033\\
\end{tabular}
\end{ruledtabular}
\end{table}
In our study of ordering in the ternary \AlInN, we observe qualitatively similar
behavior as in the case of \GaInN (compare Figs.~\ref{fig:SRO_GaInN} and
\ref{fig:SRO_AlInN}). Again, the same alternating pattern of signs for order
parameters is observed, i.e., \sro{1}{AlIn} and \sro{4}{AlIn} are negative,
whereas \sro{2}{AlIn} and \sro{3}{AlIn} are positive. Similarly to the previous
section, the SRO parameter values and concentrations for extremum ordering
\sro{i}{ext} and \xexti{i} are determined. The extremal values of order
parameters were in the case of \AlInN larger than in the case of \GaInN roughly
by a factor of 50\%, which correlates with larger lattice mismatch between AlN
and InN than between the GaN and InN. For detailed comparison, see
Table~\ref{tab:AlInNExtrema} and the analogous data for \GaInN gathered in
Table~\ref{tab:GaInNExtrema}. Important difference in the case of \AlInN is that
the composition dependencies of \sro{2}{AlIn}, \sro{3}{AlIn}, and \sro{4}{AlIn}
exhibit features of bimodal shape, particularly in lower temperatures. 
Therefore, in these cases the third-order polynomial fits carried out in order
to locate extrema, were performed in the narrowed range: $x \in [0.0,0.6]$
for \sro{2}{AlIn}, \sro{3}{AlIn} and $x \in [0.0,0.7]$ for \sro{4}{AlIn}. The
bimodal character is particularly pronounced for \sro{2}{AlIn}, \sro{3}{AlIn} in
the lowest temperature $T=873$~K. In these cases, the second maxima are clearly
visible and could be easily calculated from polynomial fits. The detailed
information about all extrema is provided in Table~\ref{tab:AlInNExtrema}. Apart
from the bimodal character, other features observed for the SRO parameters in
\AlInN are similar to \GaInN. The shape of extremum for \sro{1}{AlIn} and
\sro{2}{AlIn} is asymmetric and located in the high indium concentration range
around 70\%, a little bit higher than for the \GaInN case. For the highest
coordination shells, the extremum gets more symmetric and shifts toward 50\%
composition. Again for this alloy, the location of extremum concentration weakly
depends on temperature and changes maximally around 3\%, as presented in
Table~\ref{tab:AlInNExtrema}.

Since the properties of \AlInN lattice matched to GaN are of high practical
importance, we gather the information about the short-range order parameters of
this material in Table~\ref{tab:AlInNMatchedGaN}. In the context of wurtzite
\AlInN, usually the aluminum concentration $x\approx0.82$ is quoted as
corresponding to lattice matching. \cite{Butte2007} In the case of cubic
materials and lattice parameters employed in this study, the Vegard's law leads
to the concentration $x\approx0.79$.  For simplicity, we neglect here the small
unphysical lattice bowing effect which is a feature of the Keating model.
\cite{Lopuszynski2010} The composition of the \AlInN lattice matching GaN, is
also marked for convenience in Fig.~\ref{fig:SRO_AlInN}. It turns out that the
magnitude of SRO parameters in this case does not exceed 0.05. Interestingly,
it is the lowest for the first coordination shell, which should have the
greatest impact on the alloy properties. When it comes to modeling
implications, the situation here is analogous to \GaInN. Our calculations
indicate that for electronic structure modeling the configuration with 
\noindent SRO sign pattern $-/+/+/-$ should be the most interesting.  The
deviation from random uncorrelated alloy is the biggest for high indium
concentration around 70\%, therefore, in this region models assuming SRO=0
should have the largest systematic inaccuracy.  

\begin{table}[!htb]
\caption{ Summary of the short-range order parameters for \AlInN lattice 
          matched to GaN, i.e., for aluminum concentration $x\approx0.79$.
          \label{tab:AlInNMatchedGaN}}
\smallskip
\begin{ruledtabular}
\begin{tabular}{cdd}
\dummyBtm   &  \multicolumn{1}{c}{T=873 K}
             & \multicolumn{1}{c}{T=1673 K} \\
\hline
\dummyBig$\sro{1}{AlIn}$ & -0.014 &     -0.012      \\
\dummySml$\sro{2}{AlIn}$ &  0.049 &      0.019      \\
\dummySml$\sro{3}{AlIn}$ &  0.041 &      0.016      \\
\dummySml$\sro{4}{AlIn}$ & -0.038 &     -0.027      \\
\end{tabular}
\end{ruledtabular}
\end{table}
\subsection{Ordering in \AlGaN}

Finally, the last possible ternary combination for the examined family of
nitrides is \AlGaN. This material is very well suited for high electron
mobility transistors, operating in high-power range as well as ultraviolet
light emitters and detectors. It is also a promising building block for
biosensors.  Since the lattice mismatch and the differences in force field
parameters are the smallest in this case, also the ordering effects here are
the weakest. For this material, we carried out the studies only for two
temperatures 873 and 1673~K. None of the examined $\sro{i}{AlGa}$ values
exceeded 0.01.  Therefore, our conclusion is that \AlGaN follows quite closely
the picture of of uncorrelated random alloy, i.e., with both SRO=0 and LRO=0.

\section{Ordering in ternary alloys on the substrate
    \label{sec:OrderTernSubs}}

In the previous section we have examined ordering phenomena in bulk crystals.
However, for applications in optoelectronic devices or sensors usually thin
epitaxial layers grown on substrate are employed. In this part, we describe,
how the presence of substrate associated with biaxial strain influences the
ordering phenomena. We focus on the technologically most important case of
\GaInN and examine its behavior on a variety of substrates. However, before
moving on to the discussion of ordering, the basic facts about thin epitaxial
layers are summarized below.

\begin{figure}[!htb]
    \centering
    \includegraphics[width=0.47\textwidth]
    {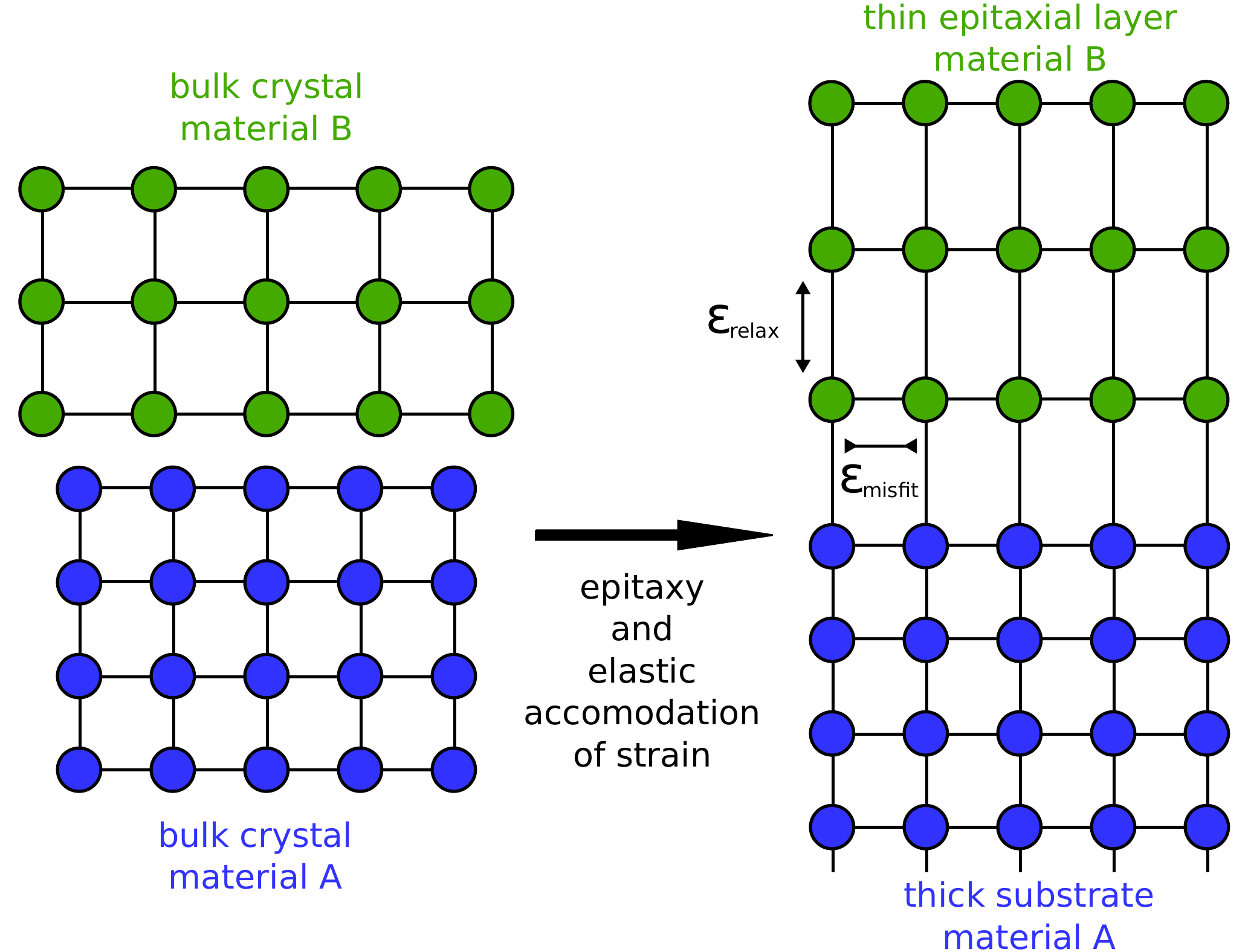}
    \caption{\label{fig:Epitaxy} (Color online) 
    The epitaxial layer in the presence of lattice mismatch between the
    layer material and the substrate material. The case of purely
    elastic accommodation of strain is presented, i.e., without misfit 
    dislocations (lattice-coherent case).}
\end{figure}

The main parameter for epitaxial layer grown on the substrate is misfit
strain. For a cubic material, it is defined as
\begin{equation}
    \epsilon_{\text{misfit}}=
        \frac{a_{\text{substrate}}-a_{\text{layer}}}{a_{\text{layer}}},
\end{equation}
where $a_{\text{substrate}}$ and $a_{\text{layer}}$ denote substrate and layer
alloy lattice constants respectively. If we consider the simplest model, the
misfit strain is compensated by purely elastic deformation of the thin film
 material. The thick substrate is assumed to stay undeformed in this approach.
 See Fig.~\ref{fig:Epitaxy} for illustration of such a situation. Within this
 model, the deformation of layer unit cell on the interface (in order to match
 the substrate lattice) is compensated by relaxation in the perpendicular
 direction. One can approximate the relaxation strain $\epsilon_{\text{relax}}$
 associated with accommodation of misfit strain using linear elasticity theory
\begin{equation}
    \epsilon_{\text{relax}}= -2 \; \frac{c_{12}}{c_{11}} \;
    \epsilon_{\text{misfit}}.
\label{eq:RelaxationStrain}
\end{equation}
However, for large misfit strains going beyond the applicability range of the
linear elasticity (few percent deformations), this expression can be a rather
rough approximation.
It is important to note that in the case of lattice mismatch between the
substrate and epitaxial,  material there exists a critical thickness $h$, beyond
which layer starts to relax the strain in the form of misfit dislocations and
other defects, instead of purely elastic deformation. This means that because of
the high defects density, epitaxial layers beyond $h$ are virtually useless from
the point of view of light-emitting heterostructures. However, estimation of
the critical thickness is a difficult problem. Various models were proposed for
this purpose. In this paper, in order to get a rough estimate of the critical
thickness, we employ a simple approach developed by Matthews and Blakeslee
\cite{Matthews1974} and later on employed to cubic nitrides. \cite{Sherwin1991}
It predicts that $h$ is given by the transcendental equation
\begin{equation}
h = \frac{a_{\text{substrate}} \; (1-\nu/4)}
       {4\sqrt{2} \pi \; (1+\nu)\; |\epsilon_{\text{misfit}}|}
    \left [ 
        \ln \left ( \frac{\sqrt{2}\;h}{a_{\text{substrate}}} \right ) +1
    \right ],
    \label{eq:CriticalThickness}
\end{equation}
where $\nu$ is the Poisson ratio of the layer material
$\nu=c_{12}/(c_{11}+c_{12})$.  Note that since both misfit strain and Poisson
ratio depend on alloy composition, the resulting critical thickness also
depends on the proportion of compounds in alloy epitaxial layer.  Employed
values of lattice constants are given in Table~\ref{tab:SubstratesForGaInN},
the composition dependencies of the elastic parameters in
Eq.~(\ref{eq:CriticalThickness}) were taken from our previous work
\cite{Lopuszynski2010} and are given by
\begin{eqnarray}
c_{11}(x)&=& 182.23 + 105.70\;x -15.44\;x\,(1 - x), \\
c_{12}(x)&=& 104.78 + \phantom{1}25.02\;x -\phantom{1}1.00\;x\,(1 - x).
\nonumber
\end{eqnarray}
For certain compositions related to large misfit strains, the real-valued
solutions of the Eq.~(\ref{eq:CriticalThickness}) do not exist, which indicates
that fabricating even very thin layers of high crystalline quality is impossible
for certain alloy/substrate combinations. However, one has to bear in mind, that
accurate modeling of the critical thickness is complicated task in itself. The
Matthews and Blakeslee approach is simple, but crude model. Therefore, it gives
more qualitative than quantitative picture of the $h$ dependence on composition
and limits of the elastic strain accommodation regime.

In the present section we examine the \GaInN ternary alloy grown on a large
variety of possible substrates, both employed experimentally as well as
possibly suitable in the future, including \mbox{3C-SiC}, \mbox{zb-ZnO}, CaO,
AlN, GaN, and InN.  Currently, the most promising well lattice matched
substrates for the zinc blende nitrides seem to be 3C-SiC or GaN obtained
either in the form of thick buffer layer on a different substrate or recently
fabricated zb-GaN free-standing crystals.
\cite{Novikov2008,Novikov2010,Novikov2010a} However, with technological
progress, other options can gain significance in the future. The information
about misfits bounds for each of these materials is gathered in
Table~\ref{tab:SubstratesForGaInN}.

\begin{table}[!t]
\caption{Misfit strain range for various substrate applicable to \GaInN alloys.
         The extremum values for pure GaN and pure InN are provided.
         \label{tab:SubstratesForGaInN}}
\smallskip

\begin{ruledtabular}
\begin{tabular}{ccdd}
     Substrate     & Lattice  
                   &\multicolumn{2}{l}{Misfit strain range for \GaInN} \\
                   & constant
                   &\multicolumn{1}{c}{$x=1$}
                   &\multicolumn{1}{c}{$x=0$}     \\
                   &
                   &\multicolumn{1}{c}{(pure GaN)}        
                   &\multicolumn{1}{c}{(pure InN)}  \dummyBtm\\
\hline
\dummyBig        3C-SiC  & 4.360 &      -3.2~\% &      -12.8~\% \\
\dummySml           AlN  & 4.374 &      -2.9~\% &      -12.5~\% \\
\dummySml           GaN  & 4.503 &       0.0~\% &       -9.9~\% \\
\dummySml         zb-ZnO & 4.580 &       1.7~\% &       -8.4~\% \\
\dummySml           CaO  & 4.811 &       6.8~\% &       -3.8~\% \\
\dummySml           InN  & 5.000 &      11.0~\% &        0.0~\% \\
\end{tabular}
\end{ruledtabular}
\end{table}

To simplify the discussion, we separate our results into two groups of
substrates related to moderate and large misfit strains, since the observed
phenomena in both cases are quite different. The first group consists of GaN,
\mbox{zb-ZnO}, and CaO, whereas the second comprises \mbox{3C-SiC}, AlN, and
InN. All the results have been obtained in $T=873$~K, corresponding to a
typical growth temperature range.

\subsection{Moderate misfits regime}

For substrates related to moderate misfit strains (i.e., GaN, \mbox{zb-ZnO},
and CaO), no long-range ordering is observed in \GaInN, similarly to the bulk
case.  Since the intensity of ordering effects in materials is quantified by
the absolute value of short-range order parameter $|\sro{i}{GaIn}|$, it is
useful to analyze the influence of the substrate in terms of $\Delta
\sro{i}{GaIn}$ defined as
\begin{equation}
\Delta \sro{i}{GaIn} = |\sro{i}{GaIn}(\text{on substrate})| - 
                       |\sro{i}{GaIn} (\text{bulk})|.
\end{equation}
If the presence of the substrate does not change the sign of $\sro{i}{GaIn}$
(which is the case for examined GaN, \mbox{zb-ZnO}, and CaO), the
interpretation of $\Delta \sro{i}{GaIn}$ is straightforward. When $\Delta
\sro{i}{GaIn}>0$, the ordering effects are increased, whereas for $\Delta
\sro{i}{GaIn}<0$, the grown thin layer behaves closer to the uncorrelated 
random alloy than the bulk case.

\begin{figure}[H]
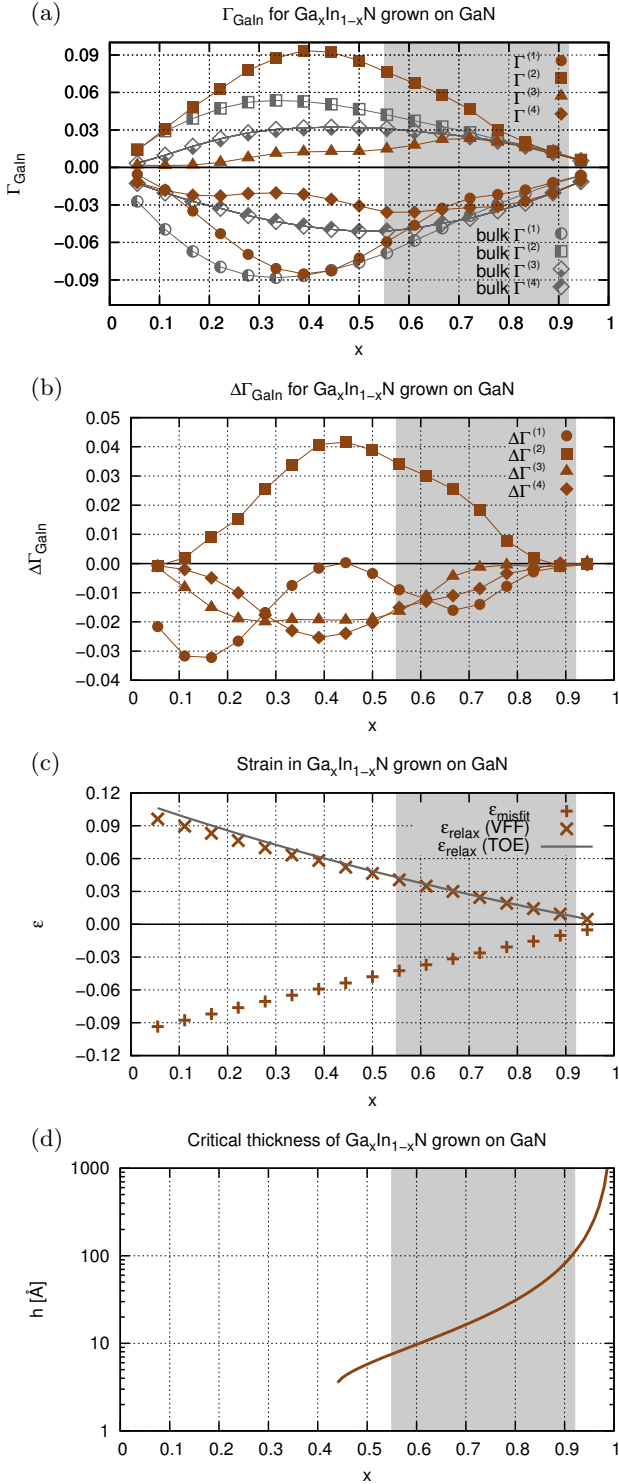

    \begin{lpic}{oto_GaInN_onGaN(8.5cm)}
    \lbl[l]{7,83;(a)} 
    \end{lpic}
    \begin{lpic}{del_GaInN_onGaN(8.5cm)}
    \lbl[l]{7,83;(b)} 
    \end{lpic}
    \begin{lpic}{lat_GaInN_onGaN(8.5cm)}
    \lbl[l]{7,83;(c)} 
    \end{lpic}
    \begin{lpic}{cri_GaInN_onGaN(8.5cm)}
    \lbl[l]{7,83;(d)} 
    \end{lpic}
    \caption{(Color online) Summary of quantities related to ordering in \GaInN
         grown on GaN. The SRO parameters \sro{i}{GaIn} (a), their difference
         with respect to bulk case $\Delta \sro{i}{GaIn}$ 
         (b), the misfit/relaxation strains calculated on the basis of 
         elasticity theory (TOE) or Keating VFF (c), and  
         the critical thickness (d). The composition range
         approximately corresponding to violet-blue-green wavelength 
         of emitted light (400--570 nm) is gray shaded.
         \label{fig:SRO_GaInNonGaN}}
\end{figure}

As an example of obtained results, the case of GaN substrate is presented in 
Fig.~\ref{fig:SRO_GaInNonGaN}. The findings for CaO and ZnO are qualitatively
similar. The graphs illustrate the comparison of the short-range order
parameters \sro{1}{GaIn}, \sro{2}{GaIn}, \sro{3}{GaIn}, and \sro{4}{GaIn}
between strained and unstrained cases. Order parameters are accompanied by the
corresponding $\Delta \sro{i}{GaIn}$ values. Also, the misfit strain
$\epsilon_{\text{misfit}}$ and the relaxation strain $\epsilon_{\text{relax}}$
dependence on composition are provided for completeness. Finally, the
information about the critical thickness $h$ calculated from the
Matthews-Blakeslee model given by Eq.~(\ref{eq:CriticalThickness}) is included.
This provides insight, as to range of composition is of practical importance.

The $\epsilon_{\text{relax}}$ presented in Fig.~\ref{fig:SRO_GaInNonGaN} (c) has
been calculated using two methods: first, within the theory of elasticity
(TOE) and Eq.~(\ref{eq:RelaxationStrain}), second by relaxing the unit cell to
the shape corresponding to minimum energy employing the Keating valence force
field (VFF). For GaN and other examined substrates, both methods give similar
results. 
The small differences emerged for compositions related to larger misfit
strains. This is intuitively well understood, since the Keating model was
constructed to recover the predictions of the TOE for small strains.
\cite{Keating1966} We also observed that for negative misfit strains
(compression in the substrate plane), the theory of elasticity slightly
overestimates the $\epsilon_{\text{relax}}$, whereas for positive misfits (in
the case of \mbox{zb-ZnO} and CaO substrate), the $\epsilon_{\text{relax}}$
predicted by 
\noindent 
the Keating force field is a little bit larger than calculated
from elasticity. When it comes to the analysis of the substrate influence on
order parameters, it can be clearly seen from
Figs.~\ref{fig:SRO_GaInNonGaN}~(a) and (b), that the most sensitive to the
presence of substrate is \sro{2}{GaIn}, as the magnitude of
$\Delta\sro{2}{GaIn}$ is the largest. Similar behavior is observed not only for
the  GaN, but also for \mbox{zb-ZnO}, and CaO. 

Interestingly, for both negative and positive misfits (present in the
case of \mbox{zb-ZnO} and CaO), the $\Delta \sro{2}{GaIn}$ is positive,
indicating that the presence of the substrate-related strain increases the
ordering within this shell. On the other hand, the quantities $\Delta
\sro{1}{GaIn}$ and $\Delta \sro{4}{GaIn}$ are mostly negative, showing the
decreasing degree of ordering compared to the bulk case for these coordination
shells. The magnitude of this effect is, however, lower than in the case of
$\Delta \sro{2}{GaIn}$. Finally, the observed behavior of $\Delta
\sro{3}{GaIn}$ is dependent on the sign of misfit strain.  For the negative
$\epsilon_{\text{misfit}}$, the decreased ordering is observed ($\Delta
\sro{3}{GaIn}<0$), and for positive values of the $\epsilon_{\text{misfit}}$,
the ordering increases ($\Delta \sro{3}{GaIn}>0$).

Let us also mention that the previous calculations by Liu and Zunger
using epitaxial cluster expansion\cite{Liu2008,Liu2009} predict that already
the influence of the GaN substrate triggers the phase separation and long-range
ordering in $T=873$~K. They obtained the full miscibility in temperatures above 
$T_M=1080$~K. Our model predicts similar phenomena, however, in order to
observe them substrates related to higher strains are necessary, as discussed
in the next section.

\subsection{Large misfits regime}

In this section, we describe the behavior of \GaInN on substrates that are
related to larger strains, i.e., \mbox{3C-SiC}, AlN, and InN. It turns out that
in these cases the long-range ordering can be triggered. This is detected by
analyzing the \mbox{sim-LRO} parameter $\lro{Ga}$ and its spread $\Delta
\lro{Ga}$, as described in Sec.~\ref{sec:CompOrder}. For conditions where
either $\lro{Ga}$ or $\Delta \lro{Ga}$ are considerably larger than in the bulk
case, various ordered structures can be observed in the configurations
generated during the Monte Carlo run. The details of formed patterns depend on
composition as well as on the magnitude and sign of strain induced by the
substrate. For the InN base layer, the \GaInN alloy undergoes tensile strain,
whereas for the \mbox{3C-SiC} and AlN the strain is compressive. Generally,
both \mbox{3C-SiC} and AlN correspond to similar misfits as displayed in
Table~\ref{tab:SubstratesForGaInN}; therefore, only the \mbox{3C-SiC} case will
be presented in detail. The effect of the AlN substrate is analogous.

\begin{figure}[H]
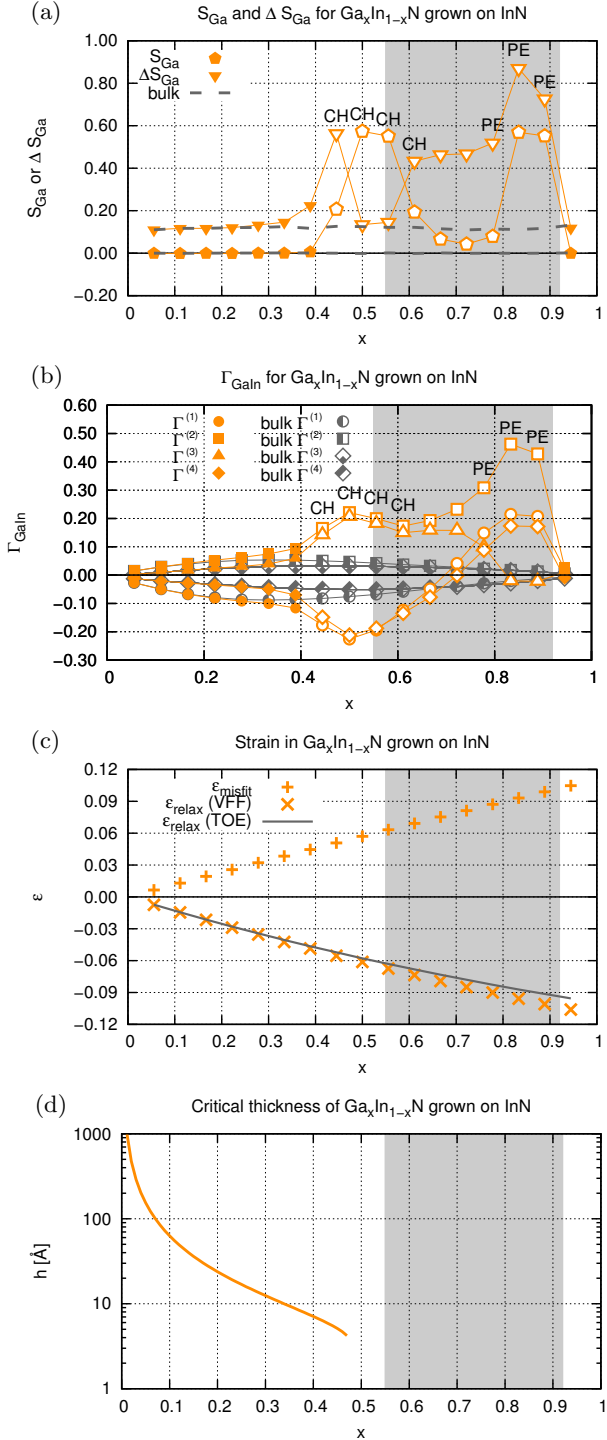

\centering
    \begin{lpic}{lro_GaInN_onInN(8.25cm)}
    \lbl[l]{7,83;(a)} 
    \end{lpic}
    \begin{lpic}{oto_GaInN_onInN(8.25cm)}
    \lbl[l]{7,83;(b)} 
    \end{lpic}
    \begin{lpic}{lat_GaInN_onInN(8.25cm)}
    \lbl[l]{7,83;(c)} 
    \end{lpic}
    \begin{lpic}{cri_GaInN_onInN(8.25cm)}
    \lbl[l]{7,83;(d)} 
    \end{lpic}
    \caption{(Color online) Summary of the quantities related to the ordering in
         \GaInN grown on InN. The \mbox{sim-LRO} parameter \lro{Ga} and its
         spread $\Delta \lro{Ga}$ (a), the SRO parameters \sro{i}{GaIn} (b), the
         misfit/relaxation strains calculated on the basis of the elasticity
         theory (TOE) or the Keating VFF (c), and the critical thickness (d).
         Open symbols denote the occurrences of the LRO. CH stands for the
         chalcopyrite and PE for the perpendicular planes ordering . The
         composition range approximately corresponding to violet-blue-green
         wavelength of emitted light (400--570 nm) is gray shaded.
         \label{fig:SRO_GaInNonInN}}
\end{figure}

\begin{figure}[!t]
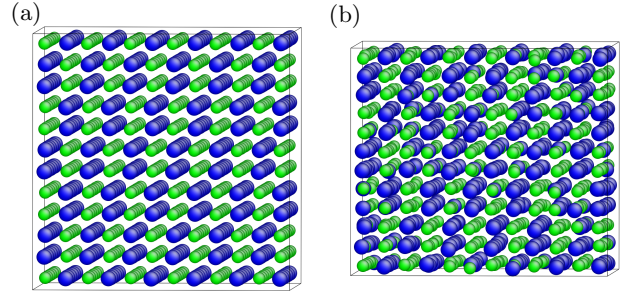

    \centering
    \begin{minipage}[c]{0.23\textwidth}
    \begin{lpic}{GaInN_chalcopyryte(3.8cm)}
    \lbl[l]{-20,385;(a)} 
    \end{lpic}
    \end{minipage}
    \begin{minipage}[c]{0.23\textwidth}
    \centering
    \begin{lpic}{GaInN_onInN_chalcopyryte(3.8cm)}
    \lbl[l]{-20,365;(b)} 
    \end{lpic}
    \end{minipage}
    \caption{(Color online) Atomic arrangement in the chalcopyrite structure
    (the CH pattern). The ideal structure (a) is compared with the sample atomic
    arrangement obtained for \GaInNcon{0.5}{0.5} on InN during Monte Carlo
    simulation (b). Note that expansion in the basal (substrate) plane and
    corresponding compression in the perpendicular plane is visible,
    together with thermal displacements.
    \label{fig:CHpattern}}
\end{figure}

\begin{figure}[!t]
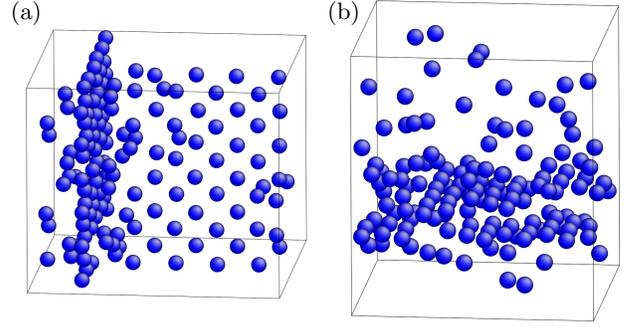

    \begin{minipage}[c]{0.23\textwidth}
    \centering
    \begin{lpic}{GaInN_onInN_planes(3.8cm)}
    \lbl[l]{-20,390;(a)} 
    \end{lpic}
    \end{minipage}
    \begin{minipage}[c]{0.23\textwidth}
    \centering
    \begin{lpic}{GaInN_onSiC_planes(3.8cm)}
    \lbl[l]{-20,390;(b)} 
    \end{lpic}
    \end{minipage}
    \caption{(Color online) Different planar ordering observed in
             $\textrm{Ga}_{0.83} \textrm{In}_{0.17} \textrm{N}$ for two
             different substrates. For InN substrate, the ordering in the
             direction perpendicular to the substrate (the PE pattern) occurs
             (a). For SiC substrate, ordering in the planes parallel to the
             substrate (the PA pattern) is present (b). Only In atoms (minor
             component) are displayed for clarity.
             \label{fig:PAPEpatterns}}
\end{figure}

Let us begin with the analysis of the InN substrate. The complete set of
characteristic quantities for this case is presented in
Fig.~\ref{fig:SRO_GaInNonInN}. Graphs include the dependencies on composition of
the long-range order parameter \lro{Ga} and its spread $\Delta \lro{Ga}$, the
short-range order parameters $\sro{i}{GaIn}$, the relaxation together with
misfit strains, and 
\noindent critical thickness. One can easily see that the
\mbox{sim-LRO} parameter indicates the presence of the long-range ordering in
the range $0.40<x<0.90$, since either the \lro{Ga} or $\Delta \lro{Ga}$ reaches
high values, compared to the bulk case. The observed dependencies of the LRO
parameters on Monte Carlo time are in these cases similar to the sample series
presented in Fig.~\ref{fig:LRO_TimeEvolution}~(b) and (c). The presence of the
LRO in the range $0.40<x<0.90$ is also indicated 
by much higher values of the $\sro{i}{GaIn}$ compared to the bulk
case. It is worth noting that the whole violet-blue-green composition region
lies completely in the discussed LRO regime for the considered InN substrate.
The visual inspection of the structures obtained during simulations reveals
that for concentrations around $x=0.5$ the so called chalcopyrite (CH) pattern
develops in the examined crystal. The ideal chalcopyrite structure is presented
in Fig.~\ref{fig:CHpattern}~(a) and compared with one of the structures
obtained during simulations in Fig.~\ref{fig:CHpattern}~(b). 
Note that even
though both the deformation due to epitaxial strain and the effect of thermal
vibrations are visible in the Monte Carlo structure, its similarity to the
perfect chalcopyrite is clear. The concentrations corresponding to the
chalcopyrite ordering pattern are marked in Fig.~\ref{fig:SRO_GaInNonInN} with
the CH label. It is well known from the literature that this type of ordering
is energetically very favorable in the case of strained semiconductor alloys,
not only nitrides.  \cite{Wei1990,Liu2007,Liu2009a} Therefore, its appearance
in our results is consistent with the previous findings. Moreover, the
chalcopyrite pattern developed in our simulations is oriented perpendicularly
to the substrate plane (CH$_{\perp}$) as predicted in the recent work of Liu
and coworkers for large misfits.\cite{Liu2009a} The study of the remaining part
of the concentration range reveals that the CH pattern is not the only ordering
option. It turns out that different atomic arrangement occurs for the higher Ga
concentration $x>0.7$ . It consists of the In planes perpendicular to the
substrate. The concentrations corresponding to this behavior are marked by
label PE in \noindent Fig.~\ref{fig:SRO_GaInNonInN}. Sample structure with this
type of ordering is presented in Fig.~\ref{fig:PAPEpatterns}~(a). Such a
behavior can be viewed as a certain mode of phase separation. 
\begin{figure}[H]
\centering
    \begin{lpic}{lro_GaInN_onSiC(8.25cm)}
    \lbl[l]{7,83;(a)} 
    \end{lpic}
    \begin{lpic}{oto_GaInN_onSiC(8.25cm)}
    \lbl[l]{7,83;(b)} 
    \end{lpic}
    \begin{lpic}{lat_GaInN_onSiC(8.25cm)}
    \lbl[l]{7,83;(c)} 
    \end{lpic}
    \begin{lpic}{cri_GaInN_onSiC(8.25cm)}
    \lbl[l]{7,83;(d)} 
    \end{lpic} 
    \caption{(Color online) Summary of the quantities related to the ordering in
         \GaInN grown on \mbox{3C-SiC}. The \mbox{sim-LRO} parameter \lro{Ga}
         and its spread $\Delta \lro{Ga}$ (a), the SRO parameters \sro{i}{GaIn}
         (b), the misfit/relaxation strains calculated on the basis of the
         elasticity theory (TOE) or the Keating VFF (c), and the critical
         thickness (d). Open symbols denote the occurrences of the LRO  of the
         parallel planes type (PA). The composition range approximately
         corresponding to violet-blue-green wavelength of emitted 
         light (400--570 nm) is gray shaded. \label{fig:SRO_GaInNonSiC}}
\end{figure}
\noindent This is also indicated by the fact that \sro{1}{GaIn}, \sro{2}{GaIn},
and \sro{4}{GaIn} shift toward high positive values, indicating the preference
toward Ga--Ga and In--In neighboring. The structures without the CH/PE symbols
correspond to mixed ordering carrying certain features of both arrangements.
The rest of the concentration range, namely for $x<0.4$ and for $x>0.9$,
manifests no long range ordering, similarly to the case of GaN, \mbox{zb-ZnO},
and CaO substrates. It is also worth noticing that the CH ordering is observed
close to the limit of elastic accommodation of strain regime, marked by the
existence of critical thickness solutions in Eq.~(\ref{eq:CriticalThickness}).
The PE ordering in turn is observed well beyond this regime. Therefore, in
reality the PE case might be difficult to observe due to the very poor
crystalline layer quality in this misfit region.

The behavior of \GaInN on substrates corresponding to large compressive strains
also leads to the formation of the long-range ordering. The detailed
dependencies on concentration for the long-range order characteristics \lro{Ga}
and $\Delta \lro{Ga}$, the short-range order parameters $\sro{i}{GaIn}$, the
relaxation and misfit strains accompanied by the critical thickness for SiC
substrate are presented in Fig.~\ref{fig:SRO_GaInNonSiC}. The analogous results
for AlN are not included, since they were almost identical. From our
simulations it emerges that for the high gallium concentrations in the range
$0.60 < x < 0.85$ the long-range ordering can be 
observed. Even though \lro{Ga} remains close to zero, the dependence
of $\Delta \lro{Ga}$ on concentration experiences abrupt change in the ordering
region. 
This change indicates that the \lro{Ga} dependence on MC time is similar
to the sample presented in Fig.~\ref{fig:LRO_TimeEvolution}~(c). This is also
confirmed by the larger values of the short-range order parameters in this
regime. It is worth noting that all \sro{i}{GaIn} become positive, indicating
Ga--Ga and In--In neighboring preference. Visual inspection of configuration
from the region with LRO reveals that indeed atoms tend to cluster in the planes
parallel to the substrate. This is depicted in Fig.~\ref{fig:SRO_GaInNonSiC} as
the PA pattern. Such a behavior can be interpreted as a certain mode of phase
separation. The sample structure corresponding to this arrangement is displayed
in Fig.~\ref{fig:PAPEpatterns}~(b). For the \mbox{3C-SiC} substrate, this
ordering occurs partially within the region of elastic strain accommodation
marked by the existence of solutions for the critical thickness model from
Eq.~(\ref{eq:CriticalThickness}), so it should be possible to access it
experimentally. Moreover, the whole discussed area of the LRO existence lies
within the technologically important violet-blue-green composition region. For
the Ga concentrations below the $x=0.6$, which correspond to the very high
compressive strains, no significant indications of the long-range order are
observed. Interestingly, the parameters \sro{1}{GaIn}, \sro{3}{GaIn}, and
\sro{4}{GaIn} in this concentration region have absolute values lower than in
the unstrained bulk case. The \sro{2}{GaIn}, in turn, similarly to the moderate
strain case, deviates mostly from the bulk ordering and its absolute value
increases. These phenomena, however, occur well outside the elastic
accommodation of strain regime, where in reality extremely poor crystal quality
would be obtained.

\section{Summary \label{sec:Summary}}
In this paper, the ordering phenomena in ternary nitride alloys have been
examined in a great detail. We have investigated how temperature, biaxial
strain related to the presence of substrate, and the change of composition in
the whole available range could influence type and degree of ordering. Our
approach included vibrational contribution, which is not so common in the
literature, where restricting to configurational degrees of freedom only is a
frequent approximation. As a starting point bulk \GaInN, \AlInN, and \AlGaN
have been investigated. The \AlGaN has been found to follow closely
uncorrelated random alloy picture, without both SRO and LRO. For the mixtures
containing indium, no signs of long-range ordering or precipitation have been
found in the bulk case; however, a considerable degree of short-range ordering
has been observed. The short-range order parameters corresponding to the first
four coordination shells followed the $-/+/+/-$ sign pattern, which agrees with
the behavior observed in simulations for \GaInN on the basis of cluster
expansion model.\cite{Chan2010} Both materials deviate the strongest from the
random uncorrelated alloy at high indium concentrations around 65\%--70\%. This
means that theoretical methods neglecting SRO should yield the largest
systematic error in this concentration range. When it comes to resolving the In
clustering controversy reported in the
literature,\cite{Galtrey2007,Kisielowski2007,Humphreys2007,Bartel2007} our
findings could provide a certain support for the experimental results showing
uniform distribution of In atoms in \GaInN samples. However, the model of bulk
alloy might not be fully appropriate here, since examined nitride alloys are
usually employed in (opto)electronic devices in the form of strained epitaxial
layers. Therefore, the next phase of our studies has been devoted to the
influence of the substrate related biaxial strain on \GaInN layers. A number of
different substrates has been examined and classified into two groups. The
first group, associated with moderate strains, contained GaN, CaO, and
\mbox{zb-ZnO}.  In this case, in the examined temperature $T=873$~K, still only
SRO has occurred. The second and more interesting group comprised substrates
yielding larger misfits, namely AlN, \mbox{3C-SiC}, and InN. For these
substrates, the appearance of the LRO has been detected by introduced
\mbox{sim}-LRO parameter.  For compressive misfit strain (AlN, \mbox{3C-SiC}),
the parallel planes (PA) pattern has developed for the In content 15\%-40\%.
For tensile misfit strain (InN) the LRO regime has covered the region  of
10\%-60\% of In content. The chalcopyrite ordering (CH) has emerged, when the
indium concentration was around 50\%, whereas for lower values down to 10\%,
the perpendicular planes (PE) pattern has been found. The PA and the PE cases
represent certain types of precipitation. The above results shed more light on
the issue of ordering in nitrides occurring in thermodynamical equilibrium. We
believe that the knowledge presented here can facilitate further modeling,
e.g., of electronic structure in nitride alloys. It can be also be helpful in
the interpretation of experimental findings.

\begin{acknowledgements}
This research was supported by the European Union within the European Regional
Development Fund through the grant Innovative Economy 
\mbox{POIG.01.01.02-00-008/08}. Pictures of atomistic structures were generated 
using the \atomeye program. \cite{Li2003}
\end{acknowledgements}

\bibliography{biblio}

\begin{thebibliography}{10}%
\makeatletter
\providecommand \@ifxundefined [1]{%
 \ifx #1\undefined \expandafter \@firstoftwo
 \else \expandafter \@secondoftwo
\fi
}%
\providecommand \@ifnum [1]{%
 \ifnum #1\expandafter \@firstoftwo
 \else \expandafter \@secondoftwo
\fi
}%
\providecommand \enquote [1]{``#1''}%
\providecommand \bibnamefont  [1]{#1}%
\providecommand \bibfnamefont [1]{#1}%
\providecommand \citenamefont [1]{#1}%
\providecommand\href[0]{\@sanitize\@href}%
\providecommand\@href[1]{\endgroup\@@startlink{#1}\endgroup\@@href}%
\providecommand\@@href[1]{#1\@@endlink}%
\providecommand \@sanitize [0]{\begingroup\catcode`\&12\catcode`\#12\relax}%
\@ifxundefined \pdfoutput {\@firstoftwo}{%
 \@ifnum{\z@=\pdfoutput}{\@firstoftwo}{\@secondoftwo}%
}{%
 \providecommand\@@startlink[1]{\leavevmode\special{html:<a href="#1">}}%
 \providecommand\@@endlink[0]{\special{html:</a>}}%
}{%
 \providecommand\@@startlink[1]{%
  \leavevmode
  \pdfstartlink
   attr{/Border[0 0 1 ]/H/I/C[0 1 1]}%
   user{/Subtype/Link/A<</Type/Action/S/URI/URI(#1)>>}%
  \relax
 }%
 \providecommand\@@endlink[0]{\pdfendlink}%
}%
\providecommand \url  [0]{\begingroup\@sanitize \@url }%
\providecommand \@url [1]{\endgroup\@href {#1}{\urlprefix}}%
\providecommand \urlprefix [0]{URL }%
\providecommand \Eprint[0]{\href }%
\@ifxundefined \urlstyle {%
  \providecommand \doi [1]{doi:\discretionary{}{}{}#1}%
}{%
  \providecommand \doi [0]{doi:\discretionary{}{}{}\begingroup
  \urlstyle{rm}\Url }%
}%
\providecommand \doibase [0]{http://dx.doi.org/}%
\providecommand \Doi[1]{\href{\doibase#1}}%
\providecommand \bibAnnote [3]{%
  \BibitemShut{#1}%
  \begin{quotation}\noindent
    \textsc{Key:}\ #2\\\textsc{Annotation:}\ #3%
  \end{quotation}%
}%
\providecommand \bibAnnoteFile [2]{%
  \IfFileExists{#2}{\bibAnnote {#1} {#2} {\input{#2}}}{}%
}%
\providecommand \typeout [0]{\immediate \write \m@ne }%
\providecommand \selectlanguage [0]{\@gobble}%
\providecommand \bibinfo [0]{\@secondoftwo}%
\providecommand \bibfield [0]{\@secondoftwo}%
\providecommand \translation [1]{[#1]}%
\providecommand \BibitemOpen[0]{}%
\providecommand \bibitemStop [0]{}%
\providecommand \bibitemNoStop [0]{.\EOS\space}%
\providecommand \EOS [0]{\spacefactor3000\relax}%
\providecommand \BibitemShut [1]{\csname bibitem#1\endcsname}%
\bibitem{ElMasry1998}%
  \BibitemOpen
  \bibfield{author}{%
  \bibinfo {author} {\bibfnamefont{N.~A.}\ \bibnamefont{El-Masry}}, \bibinfo
  {author} {\bibfnamefont{E.~L.}\ \bibnamefont{Piner}}, \bibinfo {author}
  {\bibfnamefont{S.~X.}\ \bibnamefont{Liu}},\ and\ \bibinfo {author}
  {\bibfnamefont{S.~M.}\ \bibnamefont{Bedair}},\ }%
  \bibfield{journal}{%
  \Doi{10.1063/1.120639}{\bibinfo {journal} {Appl. Phys. Lett.}}\ }%
  \textbf{\bibinfo {volume} {72}},\ \bibinfo {pages} {40} (\bibinfo {year}
  {1998})%
  \bibAnnoteFile{NoStop}{ElMasry1998}%
\bibitem{McCluskey1998}%
  \BibitemOpen
  \bibfield{author}{%
  \bibinfo {author} {\bibfnamefont{M.~D.}\ \bibnamefont{McCluskey}}, \bibinfo
  {author} {\bibfnamefont{L.~T.}\ \bibnamefont{Romano}}, \bibinfo {author}
  {\bibfnamefont{B.~S.}\ \bibnamefont{Krusor}}, \bibinfo {author}
  {\bibfnamefont{D.~P.}\ \bibnamefont{Bour}}, \bibinfo {author}
  {\bibfnamefont{N.~M.}\ \bibnamefont{Johnson}},\ and\ \bibinfo {author}
  {\bibfnamefont{S.}~\bibnamefont{Brennan}},\ }%
  \bibfield{journal}{%
  \Doi{10.1063/1.121166}{\bibinfo {journal} {Appl. Phys. Lett.}}\ }%
  \textbf{\bibinfo {volume} {72}},\ \bibinfo {pages} {1730} (\bibinfo {year}
  {1998})%
  \bibAnnoteFile{NoStop}{McCluskey1998}%
\bibitem{Liliental2006}%
  \BibitemOpen
  \bibfield{author}{%
  \bibinfo {author} {\bibfnamefont{Z.}~\bibnamefont{Liliental-Weber}}, \bibinfo
  {author} {\bibfnamefont{D.}~\bibnamefont{Zakharov}}, \bibinfo {author}
  {\bibfnamefont{K.}~\bibnamefont{Yu}}, \bibinfo {author}
  {\bibfnamefont{J.~A.}\ \bibnamefont{III}}, \bibinfo {author}
  {\bibfnamefont{W.}~\bibnamefont{Walukiewicz}}, \bibinfo {author}
  {\bibfnamefont{E.}~\bibnamefont{Haller}}, \bibinfo {author}
  {\bibfnamefont{H.}~\bibnamefont{Lu}},\ and\ \bibinfo {author}
  {\bibfnamefont{W.}~\bibnamefont{Schaff}},\ }%
  \bibfield{journal}{%
  \Doi{10.1016/j.physb.2005.12.120}{\bibinfo {journal} {Physica B}}\ }%
  \textbf{\bibinfo {volume} {376-377}},\ \bibinfo {pages} {468} (\bibinfo
  {year} {2006}),\ \bibinfo {note} {proceedings of the 23rd International
  Conference on Defects in Semiconductors}%
  \bibAnnoteFile{NoStop}{Liliental2006}%
\bibitem{Pakula2006}%
  \BibitemOpen
  \bibfield{author}{%
  \bibinfo {author} {\bibfnamefont{K.}~\bibnamefont{Paku{\l}a}}, \bibinfo
  {author} {\bibfnamefont{J.}~\bibnamefont{Borysiuk}}, \bibinfo {author}
  {\bibfnamefont{R.}~\bibnamefont{Bo{\.z}ek}},\ and\ \bibinfo {author}
  {\bibfnamefont{J.}~\bibnamefont{Baranowski}},\ }%
  \bibfield{journal}{%
  \Doi{10.1016/j.jcrysgro.2006.08.039}{\bibinfo {journal} {J. Cryst. Growth}}\
  }%
  \textbf{\bibinfo {volume} {296}},\ \bibinfo {pages} {191} (\bibinfo {year}
  {2006})%
  \bibAnnoteFile{NoStop}{Pakula2006}%
\bibitem{Galtrey2007}%
  \BibitemOpen
  \bibfield{author}{%
  \bibinfo {author} {\bibfnamefont{M.~J.}\ \bibnamefont{Galtrey}}, \bibinfo
  {author} {\bibfnamefont{R.~A.}\ \bibnamefont{Oliver}}, \bibinfo {author}
  {\bibfnamefont{M.~J.}\ \bibnamefont{Kappers}}, \bibinfo {author}
  {\bibfnamefont{C.~J.}\ \bibnamefont{Humphreys}}, \bibinfo {author}
  {\bibfnamefont{D.~J.}\ \bibnamefont{Stokes}}, \bibinfo {author}
  {\bibfnamefont{P.~H.}\ \bibnamefont{Clifton}},\ and\ \bibinfo {author}
  {\bibfnamefont{A.}~\bibnamefont{Cerezo}},\ }%
  \bibfield{journal}{%
  \Doi{10.1063/1.2431573}{\bibinfo {journal} {Appl. Phys. Lett.}}\ }%
  \textbf{\bibinfo {volume} {90}},\ \bibinfo {eid} {061903} (\bibinfo {year}
  {2007})%
  \bibAnnoteFile{NoStop}{Galtrey2007}%
\bibitem{Ozdol2010}%
  \BibitemOpen
  \bibfield{author}{%
  \bibinfo {author} {\bibfnamefont{V.~B.}\ \bibnamefont{\"{O}zd\"{o}l}},
  \bibinfo {author} {\bibfnamefont{C.~T.}\ \bibnamefont{Koch}},\ and\ \bibinfo
  {author} {\bibfnamefont{P.~A.}\ \bibnamefont{van Aken}},\ }%
  \bibfield{journal}{%
  \Doi{10.1063/1.3476285}{\bibinfo {journal} {J. Appl. Phys.}}\ }%
  \textbf{\bibinfo {volume} {108}},\ \bibinfo {eid} {056103} (\bibinfo {year}
  {2010})%
  \bibAnnoteFile{NoStop}{Ozdol2010}%
\bibitem{Humphreys2007}%
  \BibitemOpen
  \bibfield{author}{%
  \bibinfo {author} {\bibfnamefont{C.~J.}\ \bibnamefont{Humphreys}},\ }%
  \bibfield{journal}{%
  \Doi{10.1080/14786430701342172}{\bibinfo {journal} {Philos. Mag.}}\ }%
  \textbf{\bibinfo {volume} {87}},\ \bibinfo {pages} {1971} (\bibinfo {year}
  {2007})%
  \bibAnnoteFile{NoStop}{Humphreys2007}%
\bibitem{Smeeton2003}%
  \BibitemOpen
  \bibfield{author}{%
  \bibinfo {author} {\bibfnamefont{T.~M.}\ \bibnamefont{Smeeton}}, \bibinfo
  {author} {\bibfnamefont{M.~J.}\ \bibnamefont{Kappers}}, \bibinfo {author}
  {\bibfnamefont{J.~S.}\ \bibnamefont{Barnard}}, \bibinfo {author}
  {\bibfnamefont{M.~E.}\ \bibnamefont{Vickers}},\ and\ \bibinfo {author}
  {\bibfnamefont{C.~J.}\ \bibnamefont{Humphreys}},\ }%
  \bibfield{journal}{%
  \Doi{10.1063/1.1636534}{\bibinfo {journal} {Appl. Phys. Lett.}}\ }%
  \textbf{\bibinfo {volume} {83}},\ \bibinfo {pages} {5419} (\bibinfo {year}
  {2003})%
  \bibAnnoteFile{NoStop}{Smeeton2003}%
\bibitem{Kisielowski2007}%
  \BibitemOpen
  \bibfield{author}{%
  \bibinfo {author} {\bibfnamefont{C.}~\bibnamefont{Kisielowski}}\ and\
  \bibinfo {author} {\bibfnamefont{T.~P.}\ \bibnamefont{Bartel}},\ }%
  \bibfield{journal}{%
  \Doi{10.1063/1.2783976}{\bibinfo {journal} {Appl. Phys. Lett.}}\ }%
  \textbf{\bibinfo {volume} {91}},\ \bibinfo {eid} {176101} (\bibinfo {year}
  {2007})%
  \bibAnnoteFile{NoStop}{Kisielowski2007}%
\bibitem{Bartel2007}%
  \BibitemOpen
  \bibfield{author}{%
  \bibinfo {author} {\bibfnamefont{T.~P.}\ \bibnamefont{Bartel}}, \bibinfo
  {author} {\bibfnamefont{P.}~\bibnamefont{Specht}}, \bibinfo {author}
  {\bibfnamefont{J.~C.}\ \bibnamefont{Ho}},\ and\ \bibinfo {author}
  {\bibfnamefont{C.}~\bibnamefont{Kisielowski}},\ }%
  \bibfield{journal}{%
  \Doi{10.1080/14786430601146905}{\bibinfo {journal} {Philos. Mag.}}\ }%
  \textbf{\bibinfo {volume} {87}},\ \bibinfo {pages} {1983} (\bibinfo {year}
  {2007})%
  \bibAnnoteFile{NoStop}{Bartel2007}%
\bibitem{Bellaiche1998}%
  \BibitemOpen
  \bibfield{author}{%
  \bibinfo {author} {\bibfnamefont{L.}~\bibnamefont{Bellaiche}}\ and\ \bibinfo
  {author} {\bibfnamefont{A.}~\bibnamefont{Zunger}},\ }%
  \bibfield{journal}{%
  \Doi{10.1103/PhysRevB.57.4425}{\bibinfo {journal} {Phys. Rev. B}}\ }%
  \textbf{\bibinfo {volume} {57}},\ \bibinfo {pages} {4425} (\bibinfo {year}
  {1998})%
  \bibAnnoteFile{NoStop}{Bellaiche1998}%
\bibitem{Dudiy2003}%
  \BibitemOpen
  \bibfield{author}{%
  \bibinfo {author} {\bibfnamefont{S.~V.}\ \bibnamefont{Dudiy}}\ and\ \bibinfo
  {author} {\bibfnamefont{A.}~\bibnamefont{Zunger}},\ }%
  \bibfield{journal}{%
  \Doi{10.1103/PhysRevB.68.041302}{\bibinfo {journal} {Phys. Rev. B}}\ }%
  \textbf{\bibinfo {volume} {68}},\ \bibinfo {pages} {041302} (\bibinfo {year}
  {2003})%
  \bibAnnoteFile{NoStop}{Dudiy2003}%
\bibitem{Gorczyca2009}%
  \BibitemOpen
  \bibfield{author}{%
  \bibinfo {author} {\bibfnamefont{I.}~\bibnamefont{Gorczyca}}, \bibinfo
  {author} {\bibfnamefont{T.}~\bibnamefont{Suski}}, \bibinfo {author}
  {\bibfnamefont{N.~E.}\ \bibnamefont{Christensen}},\ and\ \bibinfo {author}
  {\bibfnamefont{A.}~\bibnamefont{Svane}},\ }%
  \bibfield{journal}{%
  \bibinfo {journal} {Phys. Status Solidi C}\ }%
  \textbf{\bibinfo {volume} {6}} (\bibinfo {year} {2009}),\ \doi{\bibinfo {doi}
  {10.1002/pssc.200880890}}%
  \bibAnnoteFile{NoStop}{Gorczyca2009}%
\bibitem{Gorczyca2009a}%
  \BibitemOpen
  \bibfield{author}{%
  \bibinfo {author} {\bibfnamefont{I.}~\bibnamefont{Gorczyca}}, \bibinfo
  {author} {\bibfnamefont{S.~P.}\ \bibnamefont{{\L}epkowski}}, \bibinfo
  {author} {\bibfnamefont{T.}~\bibnamefont{Suski}}, \bibinfo {author}
  {\bibfnamefont{N.~E.}\ \bibnamefont{Christensen}},\ and\ \bibinfo {author}
  {\bibfnamefont{A.}~\bibnamefont{Svane}},\ }%
  \bibfield{journal}{%
  \Doi{10.1103/PhysRevB.80.075202}{\bibinfo {journal} {Phys. Rev. B}}\ }%
  \textbf{\bibinfo {volume} {80}},\ \bibinfo {pages} {075202} (\bibinfo {year}
  {2009})%
  \bibAnnoteFile{NoStop}{Gorczyca2009a}%
\bibitem{Gorczyca2010}%
  \BibitemOpen
  \bibfield{author}{%
  \bibinfo {author} {\bibfnamefont{I.}~\bibnamefont{Gorczyca}}, \bibinfo
  {author} {\bibfnamefont{T.}~\bibnamefont{Suski}}, \bibinfo {author}
  {\bibfnamefont{N.~E.}\ \bibnamefont{Christensen}},\ and\ \bibinfo {author}
  {\bibfnamefont{A.}~\bibnamefont{Svane}},\ }%
  \bibfield{journal}{%
  \Doi{10.1063/1.3357419}{\bibinfo {journal} {Appl. Phys. Lett.}}\ }%
  \textbf{\bibinfo {volume} {96}},\ \bibinfo {eid} {101907} (\bibinfo {year}
  {2010})%
  \bibAnnoteFile{NoStop}{Gorczyca2010}%
\bibitem{Chichibu1997}%
  \BibitemOpen
  \bibfield{author}{%
  \bibinfo {author} {\bibfnamefont{S.}~\bibnamefont{Chichibu}}, \bibinfo
  {author} {\bibfnamefont{T.}~\bibnamefont{Azuhata}}, \bibinfo {author}
  {\bibfnamefont{T.}~\bibnamefont{Sota}},\ and\ \bibinfo {author}
  {\bibfnamefont{S.}~\bibnamefont{Nakamura}},\ }%
  \bibfield{journal}{%
  \Doi{10.1063/1.119013}{\bibinfo {journal} {Appl. Phys. Lett.}}\ }%
  \textbf{\bibinfo {volume} {70}},\ \bibinfo {pages} {2822} (\bibinfo {year}
  {1997})%
  \bibAnnoteFile{NoStop}{Chichibu1997}%
\bibitem{Chichibu2006}%
  \BibitemOpen
  \bibfield{author}{%
  \bibinfo {author} {\bibfnamefont{S.~F.}\ \bibnamefont{Chichibu}}, \bibinfo
  {author} {\bibfnamefont{A.}~\bibnamefont{Uedono}}, \bibinfo {author}
  {\bibfnamefont{T.}~\bibnamefont{Onuma}}, \bibinfo {author}
  {\bibfnamefont{B.~A.}\ \bibnamefont{Haskell}}, \bibinfo {author}
  {\bibfnamefont{A.}~\bibnamefont{Chakraborty}}, \bibinfo {author}
  {\bibfnamefont{T.}~\bibnamefont{Koyama}}, \bibinfo {author}
  {\bibfnamefont{P.~T.}\ \bibnamefont{Fini}}, \bibinfo {author}
  {\bibfnamefont{S.}~\bibnamefont{Keller}}, \bibinfo {author}
  {\bibfnamefont{S.~P.}\ \bibnamefont{DenBaars}}, \bibinfo {author}
  {\bibfnamefont{J.~S.}\ \bibnamefont{Speck}}, \bibinfo {author}
  {\bibfnamefont{U.~K.}\ \bibnamefont{Mishra}}, \bibinfo {author}
  {\bibfnamefont{S.}~\bibnamefont{Nakamura}}, \bibinfo {author}
  {\bibfnamefont{S.}~\bibnamefont{Yamaguchi}}, \bibinfo {author}
  {\bibfnamefont{S.}~\bibnamefont{Kamiyama}}, \bibinfo {author}
  {\bibfnamefont{H.}~\bibnamefont{Amano}}, \bibinfo {author}
  {\bibfnamefont{I.}~\bibnamefont{Akasaki}}, \bibinfo {author}
  {\bibfnamefont{J.}~\bibnamefont{Han}},\ and\ \bibinfo {author}
  {\bibfnamefont{T.}~\bibnamefont{Sota}},\ }%
  \bibfield{journal}{%
  \Doi{10.1038/nmat1726}{\bibinfo {journal} {Nat. Mater.}}\ }%
  \textbf{\bibinfo {volume} {5}},\ \bibinfo {pages} {810} (\bibinfo {year}
  {2006})%
  \bibAnnoteFile{NoStop}{Chichibu2006}%
\bibitem{Ganchenkova2008}%
  \BibitemOpen
  \bibfield{author}{%
  \bibinfo {author} {\bibfnamefont{M.~G.}\ \bibnamefont{Ganchenkova}}, \bibinfo
  {author} {\bibfnamefont{V.~A.}\ \bibnamefont{Borodin}}, \bibinfo {author}
  {\bibfnamefont{K.}~\bibnamefont{Laaksonen}},\ and\ \bibinfo {author}
  {\bibfnamefont{R.~M.}\ \bibnamefont{Nieminen}},\ }%
  \bibfield{journal}{%
  \Doi{10.1103/PhysRevB.77.075207}{\bibinfo {journal} {Phys. Rev. B}}\ }%
  \textbf{\bibinfo {volume} {77}},\ \bibinfo {eid} {075207} (\bibinfo {year}
  {2008})%
  \bibAnnoteFile{NoStop}{Ganchenkova2008}%
\bibitem{Ho1996}%
  \BibitemOpen
  \bibfield{author}{%
  \bibinfo {author} {\bibfnamefont{I.}~\bibnamefont{Ho}}\ and\ \bibinfo
  {author} {\bibfnamefont{G.~B.}\ \bibnamefont{Stringfellow}},\ }%
  \bibfield{journal}{%
  \Doi{10.1063/1.117683}{\bibinfo {journal} {Appl. Phys. Lett.}}\ }%
  \textbf{\bibinfo {volume} {69}},\ \bibinfo {pages} {2701} (\bibinfo {year}
  {1996})%
  \bibAnnoteFile{NoStop}{Ho1996}%
\bibitem{Adhikari2004}%
  \BibitemOpen
  \bibfield{author}{%
  \bibinfo {author} {\bibfnamefont{J.}~\bibnamefont{Adhikari}}\ and\ \bibinfo
  {author} {\bibfnamefont{D.~A.}\ \bibnamefont{Kofke}},\ }%
  \bibfield{journal}{%
  \Doi{10.1063/1.1686897}{\bibinfo {journal} {J. Appl. Phys.}}\ }%
  \textbf{\bibinfo {volume} {95}},\ \bibinfo {pages} {4500} (\bibinfo {year}
  {2004})%
  \bibAnnoteFile{NoStop}{Adhikari2004}%
\bibitem{Biswas2008}%
  \BibitemOpen
  \bibfield{author}{%
  \bibinfo {author} {\bibfnamefont{K.}~\bibnamefont{Biswas}}, \bibinfo {author}
  {\bibfnamefont{A.}~\bibnamefont{Franceschetti}},\ and\ \bibinfo {author}
  {\bibfnamefont{S.}~\bibnamefont{Lany}},\ }%
  \bibfield{journal}{%
  \Doi{10.1103/PhysRevB.78.085212}{\bibinfo {journal} {Phys. Rev. B}}\ }%
  \textbf{\bibinfo {volume} {78}},\ \bibinfo {eid} {085212} (\bibinfo {year}
  {2008})%
  \bibAnnoteFile{NoStop}{Biswas2008}%
\bibitem{Teles2000}%
  \BibitemOpen
  \bibfield{author}{%
  \bibinfo {author} {\bibfnamefont{L.~K.}\ \bibnamefont{Teles}}, \bibinfo
  {author} {\bibfnamefont{J.}~\bibnamefont{Furthm\"uller}}, \bibinfo {author}
  {\bibfnamefont{L.~M.~R.}\ \bibnamefont{Scolfaro}}, \bibinfo {author}
  {\bibfnamefont{J.~R.}\ \bibnamefont{Leite}},\ and\ \bibinfo {author}
  {\bibfnamefont{F.}~\bibnamefont{Bechstedt}},\ }%
  \bibfield{journal}{%
  \Doi{10.1103/PhysRevB.62.2475}{\bibinfo {journal} {Phys. Rev. B}}\ }%
  \textbf{\bibinfo {volume} {62}},\ \bibinfo {pages} {2475} (\bibinfo {year}
  {2000})%
  \bibAnnoteFile{NoStop}{Teles2000}%
\bibitem{Teles2002}%
  \BibitemOpen
  \bibfield{author}{%
  \bibinfo {author} {\bibfnamefont{L.~K.}\ \bibnamefont{Teles}}, \bibinfo
  {author} {\bibfnamefont{L.~M.~R.}\ \bibnamefont{Scolfaro}}, \bibinfo {author}
  {\bibfnamefont{J.~R.}\ \bibnamefont{Leite}}, \bibinfo {author}
  {\bibfnamefont{J.}~\bibnamefont{Furthm{\"u}ller}},\ and\ \bibinfo {author}
  {\bibfnamefont{F.}~\bibnamefont{Bechstedt}},\ }%
  \bibfield{journal}{%
  \Doi{10.1063/1.1518136}{\bibinfo {journal} {J. Appl. Phys.}}\ }%
  \textbf{\bibinfo {volume} {92}},\ \bibinfo {pages} {7109} (\bibinfo {year}
  {2002})%
  \bibAnnoteFile{NoStop}{Teles2002}%
\bibitem{Caetano2006}%
  \BibitemOpen
  \bibfield{author}{%
  \bibinfo {author} {\bibfnamefont{C.}~\bibnamefont{Caetano}}, \bibinfo
  {author} {\bibfnamefont{L.~K.}\ \bibnamefont{Teles}}, \bibinfo {author}
  {\bibfnamefont{M.}~\bibnamefont{Marques}}, \bibinfo {author}
  {\bibfnamefont{A.}~\bibnamefont{DalPino}},\ and\ \bibinfo {author}
  {\bibfnamefont{L.~G.}\ \bibnamefont{Ferreira}},\ }%
  \bibfield{journal}{%
  \Doi{10.1103/PhysRevB.74.045215}{\bibinfo {journal} {Phys. Rev. B}}\ }%
  \textbf{\bibinfo {volume} {74}},\ \bibinfo {eid} {045215} (\bibinfo {year}
  {2006})%
  \bibAnnoteFile{NoStop}{Caetano2006}%
\bibitem{Purton2005}%
  \BibitemOpen
  \bibfield{author}{%
  \bibinfo {author} {\bibfnamefont{J.~A.}\ \bibnamefont{Purton}}, \bibinfo
  {author} {\bibfnamefont{M.~Y.}\ \bibnamefont{Lavrentiev}},\ and\ \bibinfo
  {author} {\bibfnamefont{N.~L.}\ \bibnamefont{Allan}},\ }%
  \bibfield{journal}{%
  \Doi{10.1039/B409770J}{\bibinfo {journal} {J. Mater. Chem.}}\ }%
  \textbf{\bibinfo {volume} {15}},\ \bibinfo {pages} {785} (\bibinfo {year}
  {2005})%
  \bibAnnoteFile{NoStop}{Purton2005}%
\bibitem{Ferhat2002}%
  \BibitemOpen
  \bibfield{author}{%
  \bibinfo {author} {\bibfnamefont{M.}~\bibnamefont{Ferhat}}\ and\ \bibinfo
  {author} {\bibfnamefont{F.}~\bibnamefont{Bechstedt}},\ }%
  \bibfield{journal}{%
  \Doi{10.1103/PhysRevB.65.075213}{\bibinfo {journal} {Phys. Rev. B}}\ }%
  \textbf{\bibinfo {volume} {65}},\ \bibinfo {pages} {075213} (\bibinfo {year}
  {2002})%
  \bibAnnoteFile{NoStop}{Ferhat2002}%
\bibitem{Takayama2000}%
  \BibitemOpen
  \bibfield{author}{%
  \bibinfo {author} {\bibfnamefont{T.}~\bibnamefont{Takayama}}, \bibinfo
  {author} {\bibfnamefont{M.}~\bibnamefont{Yuri}}, \bibinfo {author}
  {\bibfnamefont{K.}~\bibnamefont{Itoh}}, \bibinfo {author}
  {\bibfnamefont{T.}~\bibnamefont{Baba}},\ and\ \bibinfo {author}
  {\bibfnamefont{J.}~\bibnamefont{Harris}},\ }%
  \bibfield{journal}{%
  \Doi{10.1143/JJAP.39.5057}{\bibinfo {journal} {Jpn. J. Appl. Phys}}\ }%
  \textbf{\bibinfo {volume} {39}},\ \bibinfo {pages} {5057} (\bibinfo {year}
  {2000})%
  \bibAnnoteFile{NoStop}{Takayama2000}%
\bibitem{Karpov2004}%
  \BibitemOpen
  \bibfield{author}{%
  \bibinfo {author} {\bibfnamefont{S.~Y.}\ \bibnamefont{Karpov}}, \bibinfo
  {author} {\bibfnamefont{N.~I.}\ \bibnamefont{Podolskaya}}, \bibinfo {author}
  {\bibfnamefont{I.~A.}\ \bibnamefont{Zhmakin}},\ and\ \bibinfo {author}
  {\bibfnamefont{A.~I.}\ \bibnamefont{Zhmakin}},\ }%
  \bibfield{journal}{%
  \Doi{10.1103/PhysRevB.70.235203}{\bibinfo {journal} {Phys. Rev. B}}\ }%
  \textbf{\bibinfo {volume} {70}},\ \bibinfo {pages} {235203} (\bibinfo {year}
  {2004})%
  \bibAnnoteFile{NoStop}{Karpov2004}%
\bibitem{Adhikari2004a}%
  \BibitemOpen
  \bibfield{author}{%
  \bibinfo {author} {\bibfnamefont{J.}~\bibnamefont{Adhikari}}\ and\ \bibinfo
  {author} {\bibfnamefont{D.~A.}\ \bibnamefont{Kofke}},\ }%
  \bibfield{journal}{%
  \Doi{10.1063/1.1728317}{\bibinfo {journal} {J. Appl. Phys.}}\ }%
  \textbf{\bibinfo {volume} {95}},\ \bibinfo {pages} {6129} (\bibinfo {year}
  {2004})%
  \bibAnnoteFile{NoStop}{Adhikari2004a}%
\bibitem{Takayama2001}%
  \BibitemOpen
  \bibfield{author}{%
  \bibinfo {author} {\bibfnamefont{T.}~\bibnamefont{Takayama}}, \bibinfo
  {author} {\bibfnamefont{M.}~\bibnamefont{Yuri}}, \bibinfo {author}
  {\bibfnamefont{K.}~\bibnamefont{Itoh}},\ and\ \bibinfo {author}
  {\bibfnamefont{J.~J.~S.}\ \bibnamefont{Harris}},\ }%
  \bibfield{journal}{%
  \Doi{10.1063/1.1388170}{\bibinfo {journal} {J. Appl. Phys.}}\ }%
  \textbf{\bibinfo {volume} {90}},\ \bibinfo {pages} {2358} (\bibinfo {year}
  {2001})%
  \bibAnnoteFile{NoStop}{Takayama2001}%
\bibitem{Liu2008}%
  \BibitemOpen
  \bibfield{author}{%
  \bibinfo {author} {\bibfnamefont{J.~Z.}\ \bibnamefont{Liu}}\ and\ \bibinfo
  {author} {\bibfnamefont{A.}~\bibnamefont{Zunger}},\ }%
  \bibfield{journal}{%
  \Doi{10.1103/PhysRevB.77.205201}{\bibinfo {journal} {Phys. Rev. B}}\ }%
  \textbf{\bibinfo {volume} {77}},\ \bibinfo {eid} {205201} (\bibinfo {year}
  {2008})%
  \bibAnnoteFile{NoStop}{Liu2008}%
\bibitem{Liu2009}%
  \BibitemOpen
  \bibfield{author}{%
  \bibinfo {author} {\bibfnamefont{J.~Z.}\ \bibnamefont{Liu}}\ and\ \bibinfo
  {author} {\bibfnamefont{A.}~\bibnamefont{Zunger}},\ }%
  \bibfield{journal}{%
  \Doi{10.1088/0953-8984/21/29/295402}{\bibinfo {journal} {J. Phys.: Condens.
  Matter}}\ }%
  \textbf{\bibinfo {volume} {21}},\ \bibinfo {pages} {295402} (\bibinfo {year}
  {2009})%
  \bibAnnoteFile{NoStop}{Liu2009}%
\bibitem{Chan2010}%
  \BibitemOpen
  \bibfield{author}{%
  \bibinfo {author} {\bibfnamefont{J.~A.}\ \bibnamefont{Chan}}, \bibinfo
  {author} {\bibfnamefont{J.~Z.}\ \bibnamefont{Liu}},\ and\ \bibinfo {author}
  {\bibfnamefont{A.}~\bibnamefont{Zunger}},\ }%
  \bibfield{journal}{%
  \Doi{10.1103/PhysRevB.82.045112}{\bibinfo {journal} {Phys. Rev. B}}\ }%
  \textbf{\bibinfo {volume} {82}},\ \bibinfo {pages} {045112} (\bibinfo {year}
  {2010})%
  \bibAnnoteFile{NoStop}{Chan2010}%
\bibitem{Karpov1998}%
  \BibitemOpen
  \bibfield{author}{%
  \bibinfo {author} {\bibfnamefont{S.~Y.}\ \bibnamefont{Karpov}},\ }%
  \bibfield{journal}{%
  \bibinfo {journal} {MRS Internet J. Nitride Semicond. Res.}\ }%
  \textbf{\bibinfo {volume} {3}},\ \bibinfo {pages} {16} (\bibinfo {year}
  {1998})%
  \bibAnnoteFile{NoStop}{Karpov1998}%
\bibitem{Ziman1979}%
  \BibitemOpen
  \bibfield{author}{%
  \bibinfo {author} {\bibfnamefont{J.~M.}\ \bibnamefont{Ziman}},\ }%
  \emph{\bibinfo {title} {Models of Disorder: The Theoretical Physics of
  Homogeneously Disordered Systems}},\ \bibinfo {edition} {1st}\ ed.\ (\bibinfo
  {publisher} {Cambridge University Press},\ \bibinfo {year} {1979})%
  \bibAnnoteFile{NoStop}{Ziman1979}%
\bibitem{Ducastelle1991}%
  \BibitemOpen
  \bibfield{author}{%
  \bibinfo {author} {\bibfnamefont{F.}~\bibnamefont{Ducastelle}},\ }%
  \emph{\bibinfo {title} {Order and Phase Stability in Alloys (Cohesion and
  Structure)}}\ (\bibinfo {publisher} {North Holland},\ \bibinfo {year}
  {1991})%
  \bibAnnoteFile{NoStop}{Ducastelle1991}%
\bibitem{Klein1951}%
  \BibitemOpen
  \bibfield{author}{%
  \bibinfo {author} {\bibfnamefont{M.~J.}\ \bibnamefont{Klein}},\ }%
  \bibfield{journal}{%
  \Doi{10.1119/1.1932747}{\bibinfo {journal} {Am. J. Phys}}\ }%
  \textbf{\bibinfo {volume} {19}},\ \bibinfo {pages} {153} (\bibinfo {year}
  {1951})%
  \bibAnnoteFile{NoStop}{Klein1951}%
\bibitem{Cowley1950}%
  \BibitemOpen
  \bibfield{author}{%
  \bibinfo {author} {\bibfnamefont{J.~M.}\ \bibnamefont{Cowley}},\ }%
  \bibfield{journal}{%
  \Doi{10.1103/PhysRev.77.669}{\bibinfo {journal} {Phys. Rev.}}\ }%
  \textbf{\bibinfo {volume} {77}},\ \bibinfo {pages} {669} (\bibinfo {year}
  {1950})%
  \bibAnnoteFile{NoStop}{Cowley1950}%
\bibitem{Allen1989}%
  \BibitemOpen
  \bibfield{author}{%
  \bibinfo {author} {\bibfnamefont{M.~P.}\ \bibnamefont{Allen}}\ and\ \bibinfo
  {author} {\bibfnamefont{D.~J.}\ \bibnamefont{Tildesley}},\ }%
  \emph{\bibinfo {title} {Computer Simulation of Liquids}}\ (\bibinfo
  {publisher} {Oxford University Press, USA},\ \bibinfo {year} {1989})%
  \bibAnnoteFile{NoStop}{Allen1989}%
\bibitem{Frenkel2001}%
  \BibitemOpen
  \bibfield{author}{%
  \bibinfo {author} {\bibfnamefont{D.}~\bibnamefont{Frenkel}}\ and\ \bibinfo
  {author} {\bibfnamefont{B.}~\bibnamefont{Smit}},\ }%
  \emph{\bibinfo {title} {Understanding Molecular Simulatios}},\ \bibinfo
  {edition} {2nd}\ ed.\ (\bibinfo {publisher} {Academic Press},\ \bibinfo
  {year} {2001})%
  \bibAnnoteFile{NoStop}{Frenkel2001}%
\bibitem{Keating1966}%
  \BibitemOpen
  \bibfield{author}{%
  \bibinfo {author} {\bibfnamefont{P.~N.}\ \bibnamefont{Keating}},\ }%
  \bibfield{journal}{%
  \Doi{10.1103/PhysRev.145.637}{\bibinfo {journal} {Phys. Rev.}}\ }%
  \textbf{\bibinfo {volume} {145}},\ \bibinfo {pages} {637} (\bibinfo {year}
  {1966})%
  \bibAnnoteFile{NoStop}{Keating1966}%
\bibitem{Lopuszynski2010}%
  \BibitemOpen
  \bibfield{author}{%
  \bibinfo {author} {\bibfnamefont{M.}~\bibnamefont{{\L}opuszy{\'n}ski}}\ and\
  \bibinfo {author} {\bibfnamefont{J.~A.}\ \bibnamefont{Majewski}},\ }%
  \bibfield{journal}{%
  \Doi{10.1088/0953-8984/22/20/205801}{\bibinfo {journal} {J. Phys.: Condens.
  Matter}}\ }%
  \textbf{\bibinfo {volume} {22}},\ \bibinfo {pages} {205801} (\bibinfo {year}
  {2010})%
  \bibAnnoteFile{NoStop}{Lopuszynski2010}%
\bibitem{Note1}%
  \BibitemOpen
  \bibinfo {note} {\protect \url {http://www.gnu.org/s/gsl/}}%
  \bibAnnoteFile{NoStop}{Note1}%
\bibitem{Note2}%
  \BibitemOpen
  \bibinfo {note} {\protect \url {http://developer.amd.com/libraries/acml}}%
  \bibAnnoteFile{NoStop}{Note2}%
\bibitem{Chen2008}%
  \BibitemOpen
  \bibfield{author}{%
  \bibinfo {author} {\bibfnamefont{S.}~\bibnamefont{Chen}}, \bibinfo {author}
  {\bibfnamefont{X.~G.}\ \bibnamefont{Gong}},\ and\ \bibinfo {author}
  {\bibfnamefont{S.-H.}\ \bibnamefont{Wei}},\ }%
  \bibfield{journal}{%
  \Doi{10.1103/PhysRevB.77.073305}{\bibinfo {journal} {Phys. Rev. B}}\ }%
  \textbf{\bibinfo {volume} {77}},\ \bibinfo {eid} {073305} (\bibinfo {year}
  {2008})%
  \bibAnnoteFile{NoStop}{Chen2008}%
\bibitem{Butte2007}%
  \BibitemOpen
  \bibfield{author}{%
  \bibinfo {author} {\bibnamefont{Butt{\'e}}}, \bibinfo {author}
  {\bibfnamefont{J.-F.}\ \bibnamefont{Carlin}}, \bibinfo {author}
  {\bibfnamefont{E.}~\bibnamefont{Feltin}}, \bibinfo {author}
  {\bibfnamefont{M.}~\bibnamefont{Gonschorek}}, \bibinfo {author}
  {\bibfnamefont{S.}~\bibnamefont{Nicolay}}, \bibinfo {author}
  {\bibfnamefont{G.}~\bibnamefont{Christmann}}, \bibinfo {author}
  {\bibfnamefont{D.}~\bibnamefont{Simeonov}}, \bibinfo {author}
  {\bibfnamefont{A.}~\bibnamefont{Castiglia}}, \bibinfo {author}
  {\bibfnamefont{J.}~\bibnamefont{Dorsaz}}, \bibinfo {author}
  {\bibfnamefont{H.~J.}\ \bibnamefont{Buehlmann}}, \bibinfo {author}
  {\bibfnamefont{S.}~\bibnamefont{Christopoulos}}, \bibinfo {author}
  {\bibfnamefont{G.~B.~H.}\ \bibnamefont{von H{\"o}gersthal}}, \bibinfo
  {author} {\bibfnamefont{A.~J.~D.}\ \bibnamefont{Grundy}}, \bibinfo {author}
  {\bibfnamefont{M.}~\bibnamefont{Mosca}}, \bibinfo {author}
  {\bibfnamefont{C.}~\bibnamefont{Pinquier}}, \bibinfo {author}
  {\bibfnamefont{M.~A.}\ \bibnamefont{Py}}, \bibinfo {author}
  {\bibfnamefont{F.}~\bibnamefont{Demangeot}}, \bibinfo {author}
  {\bibfnamefont{J.}~\bibnamefont{Frandon}}, \bibinfo {author}
  {\bibfnamefont{P.~G.}\ \bibnamefont{Lagoudakis}}, \bibinfo {author}
  {\bibfnamefont{J.~J.}\ \bibnamefont{Baumberg}},\ and\ \bibinfo {author}
  {\bibfnamefont{N.}~\bibnamefont{Grandjean}},\ }%
  \bibfield{journal}{%
  \Doi{10.1088/0022-3727/40/20/S16}{\bibinfo {journal} {J. Phys. D: Appl.
  Phys.}}\ }%
  \textbf{\bibinfo {volume} {40}},\ \bibinfo {pages} {6328} (\bibinfo {year}
  {2007})%
  \bibAnnoteFile{NoStop}{Butte2007}%
\bibitem{Matthews1974}%
  \BibitemOpen
  \bibfield{author}{%
  \bibinfo {author} {\bibfnamefont{J.}~\bibnamefont{Matthews}}\ and\ \bibinfo
  {author} {\bibfnamefont{A.}~\bibnamefont{Blakeslee}},\ }%
  \bibfield{journal}{%
  \Doi{10.1016/S0022-0248(74)80055-2}{\bibinfo {journal} {J. Cryst. Growth}}\
  }%
  \textbf{\bibinfo {volume} {27}},\ \bibinfo {pages} {118} (\bibinfo {year}
  {1974})%
  \bibAnnoteFile{NoStop}{Matthews1974}%
\bibitem{Sherwin1991}%
  \BibitemOpen
  \bibfield{author}{%
  \bibinfo {author} {\bibfnamefont{M.~E.}\ \bibnamefont{Sherwin}}\ and\
  \bibinfo {author} {\bibfnamefont{T.~J.}\ \bibnamefont{Drummond}},\ }%
  \bibfield{journal}{%
  \Doi{10.1063/1.347412}{\bibinfo {journal} {J. Appl. Phys.}}\ }%
  \textbf{\bibinfo {volume} {69}},\ \bibinfo {pages} {8423} (\bibinfo {year}
  {1991})%
  \bibAnnoteFile{NoStop}{Sherwin1991}%
\bibitem{Novikov2008}%
  \BibitemOpen
  \bibfield{author}{%
  \bibinfo {author} {\bibfnamefont{S.}~\bibnamefont{Novikov}}, \bibinfo
  {author} {\bibfnamefont{N.}~\bibnamefont{Stanton}}, \bibinfo {author}
  {\bibfnamefont{R.}~\bibnamefont{Campion}}, \bibinfo {author}
  {\bibfnamefont{C.}~\bibnamefont{Foxon}},\ and\ \bibinfo {author}
  {\bibfnamefont{A.}~\bibnamefont{Kent}},\ }%
  \bibfield{journal}{%
  \Doi{10.1016/j.jcrysgro.2008.06.018}{\bibinfo {journal} {J. Cryst. Growth}}\
  }%
  \textbf{\bibinfo {volume} {310}},\ \bibinfo {pages} {3964} (\bibinfo {year}
  {2008})%
  \bibAnnoteFile{NoStop}{Novikov2008}%
\bibitem{Novikov2010}%
  \BibitemOpen
  \bibfield{author}{%
  \bibinfo {author} {\bibfnamefont{S.~V.}\ \bibnamefont{Novikov}}, \bibinfo
  {author} {\bibfnamefont{C.~T.}\ \bibnamefont{Foxon}},\ and\ \bibinfo {author}
  {\bibfnamefont{A.~J.}\ \bibnamefont{Kent}},\ }%
  \bibfield{journal}{%
  \Doi{10.1002/pssa.200983412}{\bibinfo {journal} {Phys. Status Solidi A}}\ }%
  \textbf{\bibinfo {volume} {207}},\ \bibinfo {pages} {1277} (\bibinfo {year}
  {2010})%
  \bibAnnoteFile{NoStop}{Novikov2010}%
\bibitem{Novikov2010a}%
  \BibitemOpen
  \bibfield{author}{%
  \bibinfo {author} {\bibfnamefont{S.~V.}\ \bibnamefont{Novikov}}, \bibinfo
  {author} {\bibfnamefont{N.}~\bibnamefont{Zainal}}, \bibinfo {author}
  {\bibfnamefont{A.~V.}\ \bibnamefont{Akimov}}, \bibinfo {author}
  {\bibfnamefont{C.~R.}\ \bibnamefont{Staddon}}, \bibinfo {author}
  {\bibfnamefont{A.~J.}\ \bibnamefont{Kent}},\ and\ \bibinfo {author}
  {\bibfnamefont{C.~T.}\ \bibnamefont{Foxon}},\ }%
  \bibfield{journal}{%
  \Doi{10.1116/1.3276426}{\bibinfo {journal} {J. Vac. Sci. Technol. B}}\ }%
  \textbf{\bibinfo {volume} {28}},\ \bibinfo {pages} {C3B1} (\bibinfo {year}
  {2010})%
  \bibAnnoteFile{NoStop}{Novikov2010a}%
\bibitem{Wei1990}%
  \BibitemOpen
  \bibfield{author}{%
  \bibinfo {author} {\bibfnamefont{S.-H.}\ \bibnamefont{Wei}}, \bibinfo
  {author} {\bibfnamefont{L.~G.}\ \bibnamefont{Ferreira}},\ and\ \bibinfo
  {author} {\bibfnamefont{A.}~\bibnamefont{Zunger}},\ }%
  \bibfield{journal}{%
  \Doi{10.1103/PhysRevB.41.8240}{\bibinfo {journal} {Phys. Rev. B}}\ }%
  \textbf{\bibinfo {volume} {41}},\ \bibinfo {pages} {8240} (\bibinfo {year}
  {1990})%
  \bibAnnoteFile{NoStop}{Wei1990}%
\bibitem{Liu2007}%
  \BibitemOpen
  \bibfield{author}{%
  \bibinfo {author} {\bibfnamefont{J.~Z.}\ \bibnamefont{Liu}}, \bibinfo
  {author} {\bibfnamefont{G.}~\bibnamefont{Trimarchi}},\ and\ \bibinfo {author}
  {\bibfnamefont{A.}~\bibnamefont{Zunger}},\ }%
  \bibfield{journal}{%
  \Doi{10.1103/PhysRevLett.99.145501}{\bibinfo {journal} {Phys. Rev. Lett.}}\
  }%
  \textbf{\bibinfo {volume} {99}},\ \bibinfo {eid} {145501} (\bibinfo {year}
  {2007})%
  \bibAnnoteFile{NoStop}{Liu2007}%
\bibitem{Liu2009a}%
  \BibitemOpen
  \bibfield{author}{%
  \bibinfo {author} {\bibfnamefont{J.~Z.}\ \bibnamefont{Liu}}, \bibinfo
  {author} {\bibfnamefont{G.}~\bibnamefont{Trimarchi}},\ and\ \bibinfo {author}
  {\bibfnamefont{A.}~\bibnamefont{Zunger}},\ }%
  \bibfield{journal}{%
  \Doi{10.1063/1.3200234}{\bibinfo {journal} {Appl. Phys. Lett.}}\ }%
  \textbf{\bibinfo {volume} {95}},\ \bibinfo {eid} {081901} (\bibinfo {year}
  {2009})%
  \bibAnnoteFile{NoStop}{Liu2009a}%
\bibitem{Li2003}%
  \BibitemOpen
  \bibfield{author}{%
  \bibinfo {author} {\bibfnamefont{J.}~\bibnamefont{Li}},\ }%
  \bibfield{journal}{%
  \Doi{10.1088/0965-0393/11/2/305}{\bibinfo {journal} {Modell. Simul. Mater.
  Sci. Eng.}}\ }%
  \textbf{\bibinfo {volume} {11}},\ \bibinfo {pages} {173} (\bibinfo {year}
  {2003})%
  \bibAnnoteFile{NoStop}{Li2003}%
\end{thebibliography}%

\end{document}